\documentclass[a4paper,11pt]{article}
\usepackage{jheppub}
\usepackage[T1]{fontenc}
\usepackage{float}
\usepackage{graphicx}  
\usepackage{dcolumn}   
\usepackage[english]{babel} 
\usepackage{psfrag}
\usepackage{tabularx}
\usepackage{bbm,bm}
\usepackage{appendix}
\numberwithin{equation}{section}
\usepackage{rotating}
\usepackage{epsfig}

\usepackage{hyperref}
\allowdisplaybreaks
\begin{document}
\def\be{\begin{equation}}
\def\ee{\end{equation}}
\def\bea{\begin{eqnarray}}
\def\eea{\end{eqnarray}}
\def\ba{\begin{array}}
\def\ea{\end{array}}
\def\ben{\begin{enumerate}}
\def\een{\end{enumerate}}
\def\nab{\bigtriangledown}
\def\tpi{\tilde\Phi}
\def\nnu{\nonumber}
\newcommand{\eqn}[1]{(\ref{#1})}
\def\bw{\begin{widetext}}
\def\ew{\end{widetext}}
\newcommand{\half}{{\frac{1}{2}}}
\newcommand{\vs}[1]{\vspace{#1 mm}}
\newcommand{\dsl}{\pa \kern-0.5em /} 
\def\a{\alpha}
\def\b{\beta}
\def\g{\gamma}\def\G{\Gamma}
\def\d{\delta}\def\D{\Delta}
\def\ep{\epsilon}
\def\et{\eta}
\def\z{\zeta}
\def\t{\theta}\def\T{\Theta}
\def\l{\lambda}\def\L{\Lambda}
\def\m{\mu}
\def\f{\phi}\def\F{\Phi}
\def\n{\nu}
\def\p{\psi}\def\P{\Psi}
\def\r{\rho}
\def\s{\sigma}\def\S{\Sigma}
\def\ta{\tau}
\def\x{\chi}
\def\o{\omega}\def\O{\Omega}
\def\k{u}
\def\pa {\partial}
\def\ov{\over}
\def\nn{\nonumber}
\def\ud{\underline}
\def\ct{\textcolor{red}{\it cite }}
\def\qq{\text{$Q$-$\bar{Q}$ }}



\title{Probing Pole-Skipping through Scalar Gauss-Bonnet coupling}
\author{Banashree Baishya and Kuntal Nayek}
\affiliation{Department of Physics,\\
   Indian Institute of Technology Guwahati,\\
   Guwahati 781039, India}
\emailAdd{b.banashree@iitg.ac.in}
\emailAdd{nayek.kuntal@gmail.com}

\abstract{
The holographic phenomena of pole-skipping have been studied in the presence of scalar-Gauss-Bonnet interaction in a four-dimensional Anti-de Sitter-Schwarzchild black hole background. In this paper, we initiated a novel study on understanding the response of those pole-skipping points under the application of external sources. The source is identified with the holographic dual operator of the bulk scalar field with its non-normalizable solutions. We analyze in detail the dynamics of pole-skipping points in both sound and shear channels, considering linear perturbation in bulk. In the perturbative regime, characteristic parameters for chaos, namely Lyapunov exponent and butterfly velocity, remain unchanged. However, the momentum values of the pole-skipping points in various modes are found to be affected by the scalar source. Further, the diffusion coefficient has been observed to evolve non-trivially under the application of external sources.
}


\maketitle
\flushbottom

\section{Introduction:} Chaos, at the classical level, explains various macroscopic phenomena of hydrodynamics from a microscopic viewpoint. These phenomena are local criticality, zero temperature entropy, diffusion transport, Lyapunov exponent, and butterfly velocity. At the quantum level, chaos is similarly essential to studying those phenomena \cite{Gu:2016oyy, Patel:2016wdy, Grozdanov:2018atb}. Recently, chaos in many body systems has drawn tremendous interest. The AdS/CFT correspondence \cite{Maldacena:1997re, Aharony:1999ti} gives a better insight into the relation between the physics of black holes and quantum chaos. Generally, to measure quantum chaos, the \textit{out of time-ordered correlation}(OTOC) function has been proven to be a good tool in gravitational theories \cite{Shenker:2013pqa, Shenker:2014cwa}. The chaotic behaviour of a system can be described with the four-point OTOC,
 \begin{equation}
     \langle V(t,\vec{x})W(0)V(t,\vec{x})W(0)\rangle_{\beta_0}\approx e^{\lambda_L (t-|\vec{x}|/v_B)}
 \end{equation}
where, $\lambda_L$ is the Lyapunov exponent and $v_B$ is the butterfly velocity. Recently, it has been observed that we can investigate the chaotic behaviour from the level of the two-point function only \cite{Blake:2017ris, Blake:2018leo, Blake:2019otz, Blake:2021hjj, Natsuume:2019sfp}. For a chaotic system, the two-point energy density function shows non-uniqueness around some special points in momentum space $(\omega,\,k)$. These points are - where the poles and zeros of the energy density function overlap. They are marked as the pole-skipping (PS) points. For example, the boundary two-point function is a ratio of the normalized mode to the non-normalized mode of the bulk field $\Phi$, which generally takes the form as
 $G_{R}\propto\frac{\Phi_b(\omega,k)}{\Phi_a(\omega,k)}$, At the pole-skipping point, $\Phi_b(\omega_*,k_*)=\Phi_a(\omega_*,k_*)=0$ and makes the Green's function ill-defined. The line of poles is defined by $\Phi_a(\omega_*,k_*)=0$ whereas the line of zeros is given by $\Phi_b(\omega_*,k_*)=0$. Thus the pole-skipping points are some special locations in the $\omega-k$ plane. At the above special points $(\omega_{*},k_*)$ of energy-density two-point function, one can relate the parameters of chaos as,
\begin{equation}\label{chaos}
    \omega_*=i\lambda_{L},\hspace{1cm}k_*=\frac{i\lambda_{L}}{v_B}
 \end{equation}
 where $\lambda_{L}$ and $v_B$ are the Lyapunov exponent and butterfly velocity associated with the considered chaotic system. 

 The behaviour of the energy density function is universal for maximally chaotic systems. The microscopic dynamics of various hydrodynamic quantities are deeply related to the near-horizon analysis of gravity theory. Indeed, the pole-skipping points can be identified from the ingoing bulk field near the horizon. At those special points, the bulk field leads to the multi-valued Green's function at the boundary\cite{Natsuume:2019xcy}. In simple words, there is no unique ingoing solution at the horizon for those pole-skipping points. This holographic study has been performed for various bulk theories \cite{Natsuume:2019sfp, Ahn:2019rnq, Ahn:2020bks, Blake:2021hjj, Amano:2022mlu, Sil:2020jhr, Yuan:2020fvv, Kim:2020url, Choi:2020tdj}. In \cite{Ceplak:2019ymw, Natsuume:2019xcy}, the pole-skipping points have been found for the BTZ background. They have shown the intersection of the lines of poles and zeros and the existence of two regular ingoing solutions near the horizon. The pole-skipping has been also studied with finite coupling correction \cite{Natsuume:2019vcv}, with higher curvature correction \cite{Wu:2019esr} and also in the case of zero temperature \cite{Natsuume:2020snz}. Hydrodynamics transport phenomena have been studied with the pole-skipping \cite{Grozdanov:2018kkt, Grozdanov:2019uhi, Li:2019bgc, Abbasi:2020ykq, Abbasi:2020xli}. Similar pole-skipping points have been also evaluated for the fermionic models \cite{Ceplak:2019ymw, Ceplak:2021efc}. In the above articles, we have seen some of the pole-skipping points in the $\omega-k$ plane located at $\text{Im}(\omega)$ are related to chaos and they follow the same chaos bound \cite{Maldacena:2015waa}. We have also seen that these special points describe various hydrodynamic mechanisms apart from chaos, e.g., the momentum density two-point function gives shear viscosity, diffusion modes, etc.  
 So far, the holographic pole-skipping studies have been performed with higher curvature corrections \cite{Wu:2019esr}, finite coupling corrections \cite{Natsuume:2019vcv}, with various matter fields \cite{Wang:2022mcq, Ning:2023ggs} and in different black hole backgrounds \cite{Jeong:2022luo, Yuan:2023tft, Grozdanov:2023txs, Jeong:2023rck}. In this paper, we consider a particular class of higher derivative interacting theory coupled to the scalar field and study its effect on the pole-skipping phenomena. Theories beyond standard general relativity have been widely studied in the literature \cite{Stelle:1977ry}. The low energy effective action of string theory particularly brings about a special class of higher derivative term called Gauss-Bonnet, and the effect of such term in the pole-skipping context has already been explicitly studied \cite{Wu:2019esr, Natsuume:2019vcv}, particularly in more than four dimensions. Due to the effect of these corrections, the Lyapunov exponent and butterfly velocity have been analyzed. In most of the earlier pole-skipping studies, phenomena were considered with free bulk theory in five or higher dimensions. Our goal for this paper would be to understand the effect of the GB term in four-dimensional bulk. In four dimensions, one notes that pure GB term is a topological invariant, and hence does not lead to any modifications at the level of the field equations. One of the non-trivial modifications to theories with GB term is to introduce scalar fields $\phi$ coupled to the GB term \cite{Blazquez-Salcedo:2016enn}. Our approach in the present paper would be bottom-up and purely phenomenological.
With the scalar field coupled GB theory, our wish is to investigate the simultaneous effect of both higher derivative term and scalar field interaction on the pole-skipping phenomena. The scalar field in the bulk is associated with a dual scalar operator in the three-dimensional boundary field theory.
From the boundary field theory side, we further investigate how the influence of a dual scalar source at the boundary may affect the pole-skipping phenomena which may be interpreted as a stringy effect at low energy. 
For our present purpose, we consider a phenomenological form of the coupling $\zeta(\phi)\sim \phi^p$, where $p$ is an integer. In the top-down string theory construction, a higher derivative Gauss-Bonnet term usually appears with a dilaton scalar where dilaton is coupled exponentially.
We further assume the Schwarzchild-Anti de-Sitter black hole, whose dual field theory is assumed to be maximally chaotic at a finite temperature. In this background, we have studied the pole-skipping phenomena under the influence of scalar field coupled with higher curvature term. Particularly in the sound channel, the flow and decay of energy density are shown to be affected by the aforementioned interaction. Unlike the free theory, we find decay in momentum density in the shear channel at a higher value of $p$. The diffusion coefficient has been observed to have a significant effect due to scalar-GB coupling.

We briefly mention the result and methodology of our present work as follows:
The classical scalar field in the bulk can be understood as dual to a composite scalar operator (say ${\cal O}$) in the boundary thermal field theory. Therefore, the classical solution of the scalar field in the bulk with regularity condition at the horizon is interpreted in terms of linear response theory as ${\cal O}_c \propto {\cal O}_s$. The condensate of the dual operator ${\cal O}_c $ is mapped with the normalizable mode of $\phi$, and source ${\cal O}_s$ associated with the operator is identified with the non-normalizable model of $\phi$. We will consider those masses of the scalar field for which the above mapping is unique for simplicity (No alternative quantization).  
The relation between $({\cal O}_s, {\cal O}_c)$ does depend on how the bulk filed is coupled with the background metric. In all previous studies of pole-skipping, the interaction was taken to be minimal. In this paper, we assume the existence of non-minimal GB coupling, and that will naturally modify the relation between $({\cal O}_s, {\cal O}_c)$, which we observe to have a non-trivial effect on the pole-skipping phenomena.
Particularly, such interaction will be shown to affect the liner momentum $k$ appearing in the dispersion relation. However, as we are not considering the backreaction of the interaction, we do not expect any modification in the $\omega$ value of the pole-skipping points.
As we are interested in the perturbative regime, we will not allow the scalar source to increase much. Because, for a fixed perturbative parameter (here, $\lambda$), if we allow the scalar source to be large enough, its response will also be high. Then it will backreact to the background metric.
In the shear channel, we find a similar effect on $k$. For $p>3$, we find imaginary $k$ which implies the exponential decay or growth of the corresponding density function. Here we calculate the diffusion coefficient from the lowest point. It shows that the rate of diffusion decreases with the increase of scalar source and it is always below $1/4\pi T$ for $p>3$. On the other hand, in the sound channel, we find the effect of interaction for all $p>1$ are similar. In this channel, without interaction, $k^4$ has pure real ($<0$) values. Due to interaction, it encounters an imaginary part which increases with the effect of the scalar source. As the real $k^4<0$ gives $k$ with equal real and imaginary parts indicating the energy transport and decay/growth of energy density respectively. With the effect of interaction, the real and imaginary parts of $k$ become unequal. Thus one can conclude this is a result of the variation of thermal transport due to interaction.  

 We have organized the paper as follows. In section{ \ref{sec2}}, we briefly describe our model, showing Einstein's equation and background metric. We have studied the behaviour of the background scalar field and evaluated the source and condensation values. In the same section, we have studied pole-skipping for scalar field perturbation. The metric perturbations -- shear and sound modes -- have been discussed in section { \ref{sec3}}. In section { \ref{sec_chaos}}, we have calculated the chaos-related parameters, first, from the perturbed $vv$ component of the linearised Einstein equation and then from the master equation. Finally, in the end, we have concluded our results with a brief overview of the paper.\\

\label{sec2}\section{Holographic Gravity Background:} Now, in the holographic model, as we want to study pole-skipping at finite temperatures, we need to use a black hole solution in bulk. We consider a four-dimensional Anti-de Sitter Schwarzchild black hole. Holographically, the boundary theory is a three-dimensional gauge theory. The bulk metric asymptotically gives $(3+1)$ dimensional AdS space. So, the corresponding boundary theory is a finite temperature field theory. 
Initially, we consider a black hole solution with scalar propagating in the background. Then, the associated Einstein's action in the bulk theory as,
\begin{equation}\label{e_action}
\mathcal{S}_{EH}=\int d^{4}x\sqrt{-g}\left(\kappa\mathcal{R}+\Lambda-\frac{1}{2}\left((\partial\Phi)^2+m^{2}\Phi^2\right)\right),
\end{equation}
where $\kappa=(16\pi G_N)^{-1}$ is a constant related to the four-dimensional Newton's constant with mass dimensions $2$ (here we set it to unity.). The associated field equation
\begin{equation}
\mathcal{G}_{\mu\nu}\equiv \mathcal{R}_{\mu\nu}-\frac{1}{2}\mathcal{R}g_{\mu\nu}=\frac{1}{2\kappa}\Lambda g_{\mu\nu}+T_{\m\n}^\Phi,
\end{equation}
gives the $3+1$ dimensional AdS-Schwarzchild black hole solution and scalar solution as
\begin{equation}\label{4d_bh}
ds^2 = L^2\left[-r^2f(r)dt^2+\frac{dr^2}{r^2f(r)}+r^2\left(dx^2+dy^2\right)\right],\hspace{2cm}\Phi=0,
\end{equation}
where $f(r)=1-\left(\frac{r_0}{r}\right)^3$ and $L$ is the AdS radius. The stress-tensor term is given as, 
\begin{equation}
   T_{\m\n}^\Phi=\nabla_{\m}\Phi\nabla_{\n}\Phi-\frac 1 2 g_{\m\n}\left(\nabla_{a}\Phi\nabla^{a}\Phi+m^2\Phi^2\right).
\end{equation} 
In the Einstein action, $R$ is the Ricci scalar of the background \eqref{4d_bh} and $\Lambda$ is related to the cosmological constant in four dimensions. In our case, $\Lambda=6\kappa/L^2$ and  $r$ is the radial coordinate of the black hole with the horizon radius $r_0$. The horizon radius is related to the temperature $T$ of the black hole as $4\pi T=r_0^2f'(r_0)=3r_0$, where prime denotes derivative w.r.t. $r$.
Now in the action \eqref{e_action}, we have considered two-fold perturbation to achieve desired model. First we take perturbation in the background scalar $\Phi=0+\beta\times\phi(r)$ and then we consider the perturbation in the background Lagrangian with a perturbative interaction $\frac{1}{2}\lambda\zeta(\phi)\mathcal{R}_{GB}$, where $\beta$ and $\lambda$ are arbitrary coupling constants which are very small ($\ll 1$) real numbers. $\phi(r)$ is the minimally coupled scalar field of mass $m$. $\zeta(\phi)$ is a dimensionless real scalar functional . In this present study, we have considered $\zeta(\phi)=L^p\phi^p$, $p\in\mathbb{Z}^+$. In this present discussion, we will consider $L=1$. The term $\mathcal{R}_{GB}$ is the higher-ordered Gauss-Bonnet curvature term (in $4d$), which is coupled to the scalar $\phi(r)$ through $\zeta$. The Gauss-Bonnet term can be written as,
\begin{equation*}
    \mathcal{R}_{GB}=\mathcal{R}_{\mu\nu\rho\sigma}\mathcal{R}^{\mu\nu\rho\sigma}-4\mathcal{R}_{\mu\nu}\mathcal{R}^{\mu\nu}+\mathcal{R}^2 .
\end{equation*}
With the scalar-Gauss-Bonnet interaction term, the background action takes the following form as
\begin{equation}\label{pert_action}
    \mathcal{S}=\int d^4x
\sqrt{-g}\left[\kappa\mathcal{R}+\Lambda
    -\beta^2\frac{1}{2}\left( \partial_{\mu}\phi \partial^{\mu}\phi + m^{2}\phi^2\right) - \lambda\zeta(\phi) \mathcal{R}_{GB}\right] .
\end{equation}
For $p=0$, pole-skipping has been exclusively studied previously in the five dimensions \cite{Wu:2019esr} and it has considered the back-reaction of the higher curvature on the background. In our study, we are interested in $p\neq 0$ cases and treating $\lambda$ as a perturbative parameter, our background will remain unaffected by the back-reaction of the scalar field as the background value of the scalar field is $\Phi=0$. Now taking the variation of the metric tensor in \eqref{pert_action}, we get the Einstein equation as follows
\begin{align}
\nn&(\kappa-2\lambda\nabla_\r\nabla^\r\zeta(\phi))\mathcal{G}_{\mu\nu}-\frac{1}{2}g_{\mu\nu}(\Lambda+\frac{1}{2}\lambda\zeta(\phi)\mathcal{R}_{GB})\\\nn&+\lambda\zeta(\phi)\left(\mathcal{R}_{\m}^{~\r\s\ta}\mathcal{R}_{\n\r\s\ta}-4\mathcal{R}_{\r\m}\mathcal{R}^\r_\n+\mathcal{R}\mathcal{R}_{\mu\nu}\right)
-\lambda\left(\mathcal{R}\nabla_{(\m}\nabla_{\n)}\zeta(\phi)\right.\\& \left.-4\mathcal{R}_{\r(\m}\nabla_{\n)}\nabla^\r\zeta(\phi)+2\left(g_{\m\n}\mathcal{R}_{\r\s}+\mathcal{R}_{\m(\r\s)\n}\right)\nabla^\r\nabla^\s\zeta(\phi)\right)-\beta^2 T_{\m\n}^{\phi}= 0, \label{ein_eq}
\end{align}
where $\mathcal{G}_{\m\n}$ is the Einstein tensor. The aforementioned scalar field $\phi$ is a minimally coupled scalar in the black hole background \eqref{e_action}. Initially, the value of the background scalar is zero. So the stress tensor of the scalar does not appear in the Einstein equation. So we need not to consider any backreaction from the scalar and Gauss-Bonnet perturbations. We can use the standard Klein-Gordon equation with the interaction term for the scalar $\phi$ equation. In the interaction term, the scalar couples with the second-order curvature terms.
Taking this curvature coupling into account the Klein-Gordon equation of $\phi$ becomes modified as,
\begin{equation}\label{scal_eq}
    \frac{1}{\sqrt{-g}}\partial_\mu\left(\sqrt{-g}g^{\mu\nu}\partial_\nu\phi\right)-m^2\phi+\frac{\tilde\lambda}{2}\mathcal{R}_{GB}\frac{\partial}{\partial\phi}\zeta(\phi)=0.
\end{equation}
This correction of the KG equation will give $\tilde\lambda=\lambda/\beta^2$ order additional contribution in the scalar field solution. Looking at the bulk action, as we are interested in the linear $\lambda$ interaction, the correction in the scalar field solution will not affect our results. So we can discard this correction in the KG equation in most of the cases where the full bulk solution $\phi$ is considered.


We would aim to compute the near horizon in going modes and their properties. Therefore, it is fruitful to perform our calculations in the ingoing Eddington-Finkelstein co-ordinate. So, we consider $v=t+r_*$, where $v$ is the null coordinate and $r_*$ is the tortoise coordinate. The metric \eqref{4d_bh} transforms into,
\begin{equation}\label{ef_bh}
ds^2=-r^2f(r)dv^2+2dvdr+r^2\left(dx^2+dy^2\right) .
\end{equation}
The metric \eqref{4d_bh} is singular at $r=r_0$. In this new coordinate, the apparent singularity is removed. The metric has rotational symmetry in the $(x,y)$ plane.

In this background, solving the Klein-Gordon equation near the boundary $(r\rightarrow{\infty})$ gives,
\begin{equation}\label{zz} 
\lim_{r\rightarrow\infty}\phi(r)=\mathcal{O}_{s}r^{\Delta-3}+\mathcal{O}_{c}r^{-\Delta} .
\end{equation}
 Where, at infinity (where is our boundary), the leading coefficient $\mathcal{O}_{s}$ is the source, and the subleading coefficient $\mathcal{O}_{c}$ is the condensation of the dual boundary dual operator. Since we choose the standard quantization, the scaling dimension of the dual operator is $\Delta={3}/{2}+\sqrt{{9}/{4}+m^{2}}$. There is a lower bound on the scalar mass called the bound of BF (Breitenlohner and Freedman) which states that $m^{2} \geq-{d^2}/{4}$ for $(d+1)$ gravitational background. For the scalar mass $m^2\geq-\frac{d^2}{4}$, $\mathcal{O}_c$ is always a normalizable mode, which is the standard quantization of the scalar. In this range $\mathcal{O}_s$ acts as source and $\mathcal{O}_c$ as response. But, for $-\frac{d^2}{4}\leq m^2\leq-\frac{d^2-4}{4}$, both $\mathcal{O}_s$ and $\mathcal{O}_c$ are normalizable. So they source two different theories. Besides the standard one, $\mathcal{O}_c$ sources an alternative quantization in this mass range.

We can easily get the source and condensation from equation \eqref{zz} by performing some algebra, as shown in \cite{Chakrabarti:2019gow} and get,
\begin{align}\label{osoc}
\mathcal{O}_{s}&=\lim_{r\rightarrow\infty}\frac{r^{3-\Delta}\left(\Delta\phi(r)+r\phi'(r) \right) }{2\Delta-3}, \\
\mathcal{O}_{c}&=\lim_{r\rightarrow\infty}\frac{r^{\Delta}\left((\Delta-3)\phi(r)-r\phi'(r) \right) }{2\Delta-3}.
\end{align}
Important to remember that due to regularity conditions at the horizon, we have only one free parameter, and from the boundary field theory point of view the appropriate free parameter ${\cal O}_s$ is identified as the source in the boundary which is applied on the system under consideration. Our goal is to investigate the response of such perturbation parameterized by $\mathcal{O}_s$ on the dual field theory particularly on pole-skipping phenomena. 
To this end let us further point out that in some parameter space, the real scalar field can undergo source-free condensation if the underlying black hole is charged (see \cite{Chakrabarti:2019gow}). We consider an uncharged black hole, therefore, this does not arise. We leave it for our future study.
\subsection{Stability analysis of background}
\begin{figure}[ht]
\centering
\includegraphics[width=0.45\textwidth,height=4cm]{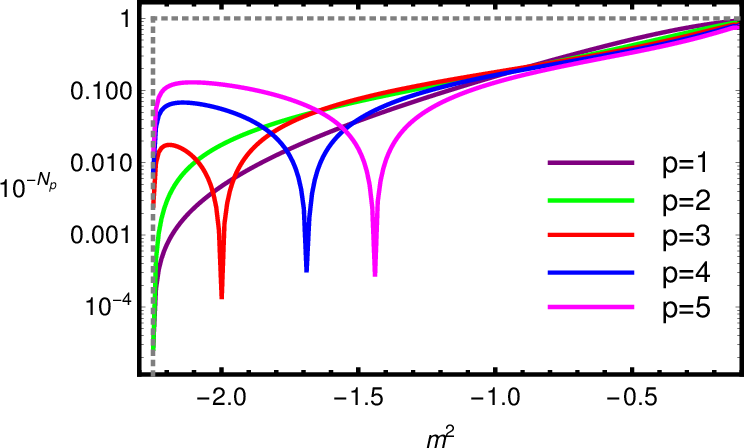}
\includegraphics[width=0.45\textwidth,height=4cm]{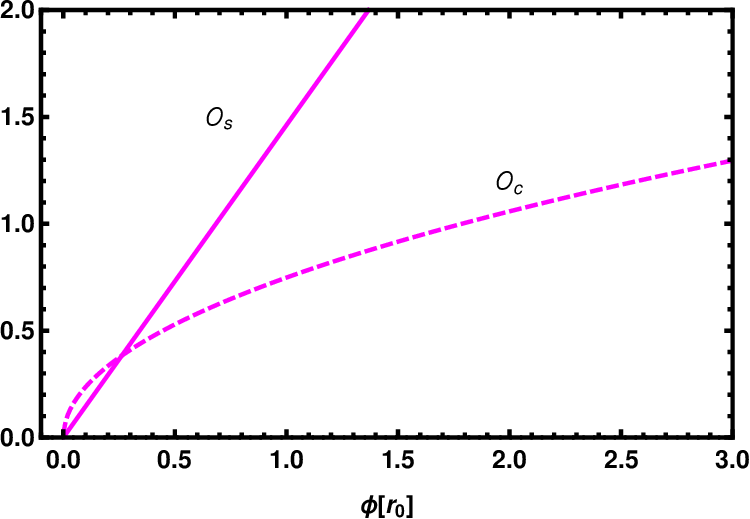}
\caption{Left: $10^{-N_p}$ vs $m^2$ for $p=1,\,2,\,3,\,4\,\&\,5$. Right: The plot of $\mathcal{O}_s$ (solid line) and $\mathcal{O}_c$ (dashed line) with $\phi(r_0)$ for $r_0=1$. Here we take scalar mass $m^2=-2$.}
\label{plot_ph0}
\end{figure}

Now we would like to comment on the stability of the Schwarzchild-AdS under the given perturbation. In terms of free energy, one expects the perturbative contribution to be sufficiently smaller than the original background contribution.
Under a given perturbative interaction, the stability of the system is broadly indicated by the thermodynamic stability. Due to perturbation, the change of free energy of the system should be negligible so that it does not disturb the thermodynamic stability. From the thermal partition function, one can show that the thermodynamic free energy $F$ is related to the on-shell bulk action $\mathcal{S}^E$ in Euclidean signature as $F=T\mathcal{S}^E$ \cite{Kim:2007qk}. So comparing the bulk on-shell one can conclude about the stability of the gravity background. So we will perform the holographic renormalization which one needs to regulate the divergence due to the asymptotic AdS. We first consider that the asymptotic boundary is located at $r=r_b$ and add the appropriate counter term $S_{ct}$ \cite{Bianchi:2001kw, Bianchi:2001de,Emparan:1999pm,Julie:2020vov}. 
\begin{align}
    \nn S_{ct}=\begin{cases}&\int d^3x\times r_b^3(2\kappa-\frac{3-\Delta}{2}\phi^2(r_b)-\frac{12\lambda}{(\Delta-3)p+3}\phi^p(r_b)) \quad\quad\text{for }\Delta \neq 3-\frac{3}{p}\\\\
     &\int d^3x\times r_b^3(2\kappa-\frac{3-\Delta}{2}\phi^2(r_b)-12\lambda\phi^p(r_b)\ln{r_b}) \quad\quad\text{for }\Delta=3-\frac{3}{p}.\end{cases}
\end{align}
Now, we perform integration on the radial coordinate and finally, we take $r_b\to\infty$ limit in it. Thus we get total on-shell action free from all divergences. Since our background is asymptotically AdS$_4$, the Gibbons-Hawking-York surface contribution is negligible ($\sim\frac{\Phi(r_b)^p}{r_b^3}$) with asymptotic correction. We will just calculate the change of on-shell action with respect to the unperturbed on-shell action to show the stability of the background. Given the background metric and solving scalar field $\phi$ from K-G equation (as we are interested in the linear perturbation in action, we exclude $\lambda$ term correction of K-G equation, we numerically find the quantity $N_p(m^2)=\text{Log}_{10}\left|\frac{1}{\lambda}\frac{S_{\lambda}}{S_0}\right|$, where
\begin{align*}
    S_0 &=  \int_{r_0}^{r_b} dr\sqrt{-g}\left[\kappa\mathcal{R}+\Lambda-\frac{1}{2}\left(g^{rr}(\partial_r\phi(r))^2+m^2\phi(r)^2\right)\right]+ r_b^3\left[2\kappa-\frac{3-\Delta}{2}\phi^2(r_b)\right].\\
    S_{\lambda}& =  \int_{r_0}^{r_b} dr\sqrt{-g}\left[\frac{\lambda}{2}\phi(r)^p\mathcal{R}_{GB}\right] - \frac{12\lambda}{(\Delta-3)p+3}r_b^3\phi^p(r_b) \quad\quad\text{for }\Delta \neq 3-\frac{3}{p}\\
      =&\int_{r_0}^{r_b} dr\sqrt{-g}\left[\frac{\lambda}{2}\phi(r)^p\mathcal{R}_{GB}\right] - 12\lambda r_b^3\phi^p(r_b)\ln{r_b} \quad\quad\text{for }\Delta=3-\frac{3}{p}.
\end{align*}
Now we numerically plot the quantity $N_p$ with $m^2$ in Figure 1. So to be in perturbative regime we need $\left|\frac{S_{\lambda}}{S_0}\right|=10^{N_p}\lambda\ll 1$, i.e., $0<\lambda\ll 10^{-N_p}$.  From the numerical plot, we see that as the absolute value of scalar mass decreases the range of $\lambda$ value becomes narrow to maintain a stable perturbative regime. Thus we get an upper-bound on $\lambda$, i.e. $0\leq\lambda\ll 10^{-N_p}$. This is shown in Figure \ref{plot_ph0} for various values of $p$.  For example, if $m^2=-2$ and $p=2$, we have $N_p=1.44$, therefore $0<\lambda\ll 0.035$. Similarly for $m^2=-2$ and $p=3$ we have $0<\lambda\ll 0.001$.

\subsection{Scalar field perturbation}
\begin{figure}[ht]
\centering\includegraphics[width=0.45\textwidth,height=4cm]{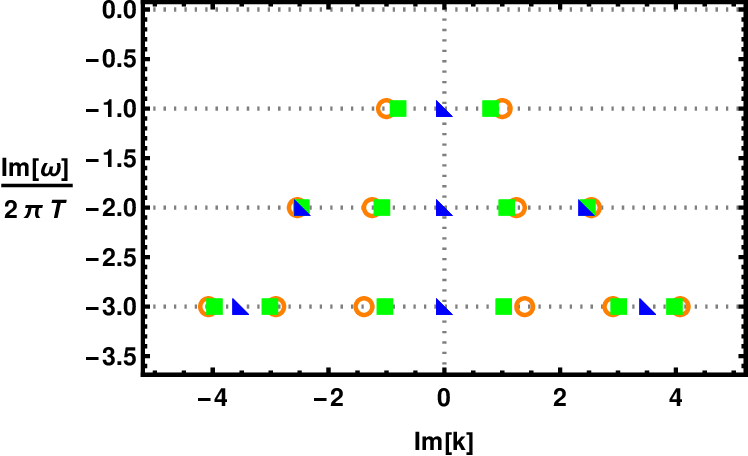}
\includegraphics[width=0.45\textwidth,height=4cm]{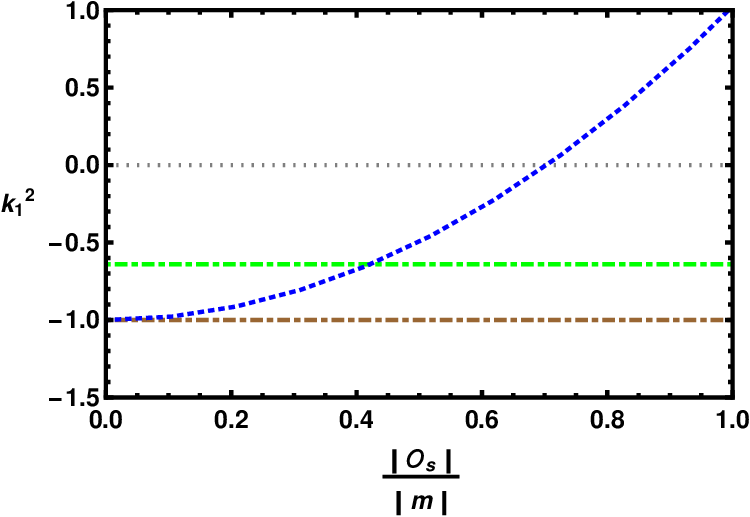}
\caption{Left: The plot of $\frac{\text{Im}\left[\omega\right]}{2\pi T}$ vs $\text{Im}\left[k\right]$ at $|\mathcal{O}_s|=5.167|m|$ for $p=1$ (orange circle), $p=2$ (green rectangle) and $p=4$ (blue triangle). Right: The plot of $k_1^2$ vs $\frac{|\mathcal{O}_s|}{|m|}$ for $p=1$ (brown dot-dashed line), $p=2$ (green dot-dashed line) and $p=4$ (blue dotted line). Here we have taken scalar mass $m^2=-2,\,\lambda=10^{-2}$ and $r_0=1$.}
\label{scal_plot}
\end{figure}

In this section, we study the effect of the perturbation on the dispersion relation associated with the scalar field $\phi$ which is a minimally coupled scalar with mass $m$. This scalar field $\phi$  is regular at the horizon and decays in the asymptotic limit. With these conditions, the solution of the scalar can be found from the equation \eqref{scal_eq}. 
Now assuming the scalar field is a function of the radial coordinate $r$ only, i.e., $\z(\phi)=\phi(r)^p$. We take the near-horizon expansion of the field as $\phi(r)=\sum_{n=0}^\infty\phi^{(n)}(r_0)(r-r_0)^n$ where, $\phi^{(n)}\equiv\frac{d^n\phi(r)}{dr^n}|_{r=r_0}$. 
From this series, the first derivative of $\phi$ at $r=r_0$ can be found as,
\begin{equation}
    3r_0\phi'\left(r_0\right)  =  m^2 \phi \left(r_0\right)-18 \lambda\, p\,\phi\left(r_0\right)^{p -1}.
\end{equation}
Similarly, we can also find the higher order derivatives in terms of $\phi(r_0)$. 
We can solve the scalar field from the K-G equation numerically by providing some horizon value to the scalar field. From this solution, we can evaluate $\mathcal{O}_s$ and $\mathcal{O}_c$ as shown in \eqref{osoc}. For the near horizon study, the regularity condition of the scalar field on the horizon is very important. So, for numerical evaluation of the source $\mathcal{O}_s$ or to get a consistent solution of $\phi(r)$; $\phi(r_0)$ should be finite and small enough so that the near-horizon expansion remains convergent. From the \figurename{ \ref{plot_ph0}} we see that source $\mathcal{O}_s$ and its near horizon value $\phi(r_0)$ is exactly linear. On the other hand, $\mathcal{O}_c$ is linear with only a small value of $\phi(r_0)$ and so with $\mathcal{O}_s$. Both of these quantities are monotonically increasing with the near horizon value of the scalar. Because of this behaviour, we will see later that the pole-skipping points show similar behaviour with sources of both of the allowed quantizations.
Now to study the dispersion relation of the scalar field, we take the perturbation $\phi(r)\to\phi(r)+e^{-i\omega v+ik x}\varphi(r)$. The linearized equation from \eqref{scal_eq} is
\begin{align}\label{pertsc}
    &r^2 f(r) \varphi''(r)+\left(r^2 f'(r)+4 r f(r)-2 i \omega \right)\varphi'(r)+ (6 \lambda  (p -1)p (f(r)^2\nn\\&-2 f(r)+3) \phi (r)^{p -2}-\frac{k^2+m^2 r^2+2 i r \omega }{r^2})\varphi(r)=0.
\end{align}
Expanding the solution near the horizon $r=r_0$ and using the matrix method as given in \cite{Ceplak:2019ymw}, we get the pole-skipping points $(\o,k)$. We find the lowest order point is $\o_1=-\frac{3}{2}ir_0=-2i\pi T$ and 
\begin{eqnarray*}
    && k_1^2+r_0^2 \left(m^2-18 \lambda p(p-1)\phi\left(r_0\right)^{p -2}+3\right)=0.
\end{eqnarray*}
Without any perturbation  ($\lambda=0$), we get the results for pure Schwarzchild black hole $k_1^2=-(3+m^2)r_0^2$, i.e., $k_1$ is completely imaginary. But, due to the effect of the interaction, $k_1$ can be real after a particular value of $\mathcal{O}_s$. Similar behaviour is also found for the higher-order pole-skipping points. Though we keep $\lambda$ small enough in the perturbative regime, $k_1^2$ becomes positive as the scalar source increases. So $k_1$ becomes real. From the equation of the perturbed scalar \eqref{pertsc}, it is clear to predict that for $p=0$ and $1$, there is no effect on $(\o,k)$, i.e., we get the values of the black hole background without any perturbation. For $p\geq 2$, $p$ effects $k_1$ in similar way as $\lambda$ does. For the small enough $\mathcal{O}_s$, we have found $k_1$ in the imaginary plane which has been plotted in the left panel of Figure \ref{scal_plot}.  
Here we have plotted first three PS points $(\o,k)$ in the complex plane for $p=1,\,2\,\&\,4$\footnote{In the Figure \ref{scal_plot}, $p=3$ case has been skipped to plot. The reason is the following. From Figure \ref{plot_ph0}, we have $\lambda\lesssim 10^{-4}$ for $m^2=-2.0$ and $p=3$. At this narrow range of $\lambda$, the shift of momentum value is visually negligible. But in this figure, our goal is to neatly represent the variation of the momentum values for various $p$. So we choose a higher $\lambda$ to get enough shift in momentum. We have taken $\lambda=10^{-2}$ which is no more stable region for $p=3$ but shows clear variation for $p=1,\,2\,\&\,4$ at that mass value.}. For $p=1\,\&\,2$, we have found $2n$-number of points for $k_n$, i.e., $n$ number of complex roots for $k_n$. However, for $p=4$, we have found one real and $n-1$ complex root of each $k_n$. Because of these real roots, we have three points on the $\text{Im}(k)$ axis. For $p=2$, the interaction imposes a constant shift in $k$. But for $p\geq 3$ the shift due to the interaction is proportional to the source. So, as the source goes to zero, $k_n$ becomes the same as the pure AdS black hole. These have been shown in the right panel of the same figure. Here, we have presented the variation of $k_1^2$ with the scalar source $\mathcal{O}_s$ for $\phi\mathcal{R}_{GB},\,\phi^2\mathcal{R}_{GB}\,\&\,\phi^4\mathcal{R}_{GB}$. It is found that $k_1$ becomes real-valued above a certain value of $\mathcal{O}_s$. Now, if we allow only the imaginary values of $k_1$, we need to put a cutoff on $\mathcal{O}_s$. The same behaviour can be found for the higher order $k$. For $\phi^3\mathcal{R}_{GB}$ interaction, we find similar behaviour of pole-skipping points as they do for $p>2$. But for this interaction, the allowed value of $\lambda$ is so small that placing it in the same graph is not convenient.
\label{sec3}\section{Metric perturbations:}
In the pole-skipping phenomena, we study the properties of the stress-energy tensor of the boundary field theory. Now with AdS/CFT duality, the bulk fields are mapped to boundary operators. Therefore, the boundary stress-energy tensors are associated with the metric perturbation of the bulk. In bulk, we consider the metric perturbation 
$
g_{\mu\nu}\to g_{\mu\nu}+e^{-i\omega v+i k x}\delta g_{\mu\nu}(r),$
where $\omega$ and $k$ are the energy and momentum parameters of the fluctuation, which propagates radially. So, in the boundary field theory, we have two point correlators which are $<T_{vv},T_{vv}>$, $<T_{vv},T_{vx}>$, $<T_{vv},T_{xx}>$, $<T_{vv},T_{yy}>$ in longitudinal mode and $<T_{vy},T_{vy}>$, $<T_{vy},T_{xy}>$, $<T_{xy},T_{xy}>$ in a transverse mode where $T_{\mu\nu}$ is the stress-energy tensor on the boundary. The metric perturbation: $\delta g_{vv},\,\delta g_{vx},\,\delta g_{xx},\,\delta g_{yy}$ and $\delta g_{vy},\,\delta g_{xy}$ are associated to the above two modes respectively. We impose the radial gauge condition $\delta g_{r\mu}=0$ for all $\mu$. We also use the traceless perturbation for simplicity, i.e., $g^{\mu\nu}\delta g_{\mu\nu}=0$ which gives $\delta g_{yy}=-\delta g_{xx}$. The longitudinal modes are the scalar modes, it does not couple with a minimally coupled scalar. Therefore we can perturb only $g_{\mu\nu}$ without effecting $\phi$. Finally, we have three independent perturbations in longitudinal mode and two in transverse mode.

\subsection{Shear Channel} As the momentum vector $(\omega,k)$ of the metric fluctuation is taken along $(v,x)$-plane, for shear mode, we consider the components coupled to $y$-direction. Here we take $g_{xy}$ and $g_{vy}$ as the only non-vanishing perturbations and these are completely decoupled from the longitudinal perturbations. These are associated with $T_{vy}$, $T_{xy}$ on the boundary. The linearised Einstein equations will give the dynamics of these fluctuations. At some special values of $(\omega,\,k)$, the solution of those equations near the horizon becomes non-unique and gives more than one independent solution. Those special points $(\omega,\,k)$ in this holographic gravity background are connected to the intersection of poles and zeros of the boundary Greens function, $G_{\mu y,\nu y}$ where $\mu,\,\nu=v,\,x$. 
Now we put these perturbations in the metric \eqref{ef_bh} and find the linearised form of the field equation with only non-vanishing perturbations $g_{xy}$ and $g_{vy}$. We find that $vy,\,ry$ and $xy$ components of the linearised equations are only non-trivial, whereas other equations are self-satisfied. Out of these three equations, we find two coupled second-order differential equations as $\delta g_{vy}''(r)=f_1\left(\delta g_{vy}',\,\delta g_{vy},\,\delta g_{xy}\right)$ and $\delta g_{xy}''(r)=f_2\left(\delta g_{vy}',\,\delta g_{vy},\,\delta g_{xy}\right)$. Again, under diffeomorphism transformation with the vector field $e^{-i \omega v+i k x}\xi^\mu$, one can show that $\delta g_{vy}$ and $\delta g_{xy}$ will form a gauge invariant combination $\mathcal{Z}_{sh}$ as,
$
\mathcal{Z}_{sh}=\left(\omega \delta g_{xy}+k \delta g_{vy} \right)/r^2
$
. So, two second-order differential equations (DE) of $\delta g_{vy}$ and $\delta g_{xy}$ combine into a single second-order DE of $\mathcal{Z}_{sh}$. The final master equation is
\begin{equation}\label{sh_master_eq}
\mathcal{M}_{sh}\mathcal{Z}''_{sh}(r)+\mathcal{P}_{sh}\mathcal{Z}'_{sh}(r)+\mathcal{Q}_{sh}\mathcal{Z}_{sh}(r)=0 .
\end{equation}
Where, the coefficients $\mathcal{M}_{sh},\,\mathcal{P}_{sh}$ and $\mathcal{Q}_{sh}$ are functions of $\omega, k$ and $\phi(r)$. The details expressions are given in the appendix. There we have considered the coefficients up to $\lambda$ order. As $\lambda= 0$ the master equation reduces to the same as the pure AdS black hole. The near horizon structure of the master variable is taken as follows.
$
    \mathcal{Z}_{sh}=\sum_{n=0}Z_n\times (r-r_0)^n
$
 . Now we expand the master equation \eqref{sh_master_eq} around $r=r_0$. At zeroth order $\mathcal{O}((r-r_0)^0)$, it gives the linear algebraic equation of $Z_0$ and $Z_1$. The coefficients of $Z_0$ and $Z_1$ are functions of two primary variables $\omega$ and $k$. So, at a particular point, $\omega=\omega_1$ the vanishing of the coefficient of $Z_1$ indicates that $Z_1$ is arbitrary. Again at the same $\omega$ value, we find a special value of $k=k_1$ where the coefficient of $Z_0$ vanishes. Therefore at the point $(\omega_1,\,k_1)$ the near horizon solution of $\mathcal{Z}_{sh}$ is defined with two arbitrary parameter $Z_0,\,Z_1$ and the solution is combination of two arbitrary solutions $C_1(r-r_0)Z_0+C_2(r-r_0)Z_1$. So we find a non-unique solution at the point $(\omega_1,\,k_1)$ -- which is the first order pole-skipping point. Here we find $\omega_1=-\frac{3}{2}ir_0$ and   
\begin{equation}\label{shear_k1}
 k_1^2 = 3 r_0^2\left[1 -3\lambda\phi(r_0)^{p}\frac{\xi(2\xi^2-\xi-17)}{2\xi+1}\right]
\end{equation}
where, $\xi=m^{2}p/3$.   
We find $\omega_1$ same as the previous result \cite{Blake:2019otz} for AdS$_4$ black hole. But $k_1$ contains a non-trivial shift due to the interaction. At $\lambda=0$, it gives the same $k_1^2$ as given in \cite{Blake:2019otz}. With nonzero $\lambda$, the shift in momentum depends on the details properties of the scalar field and its interaction namely, power $p$ of the interacting field $\phi$, the value of the field at horizon $\phi(r_0)$ and mass of it $m$. Now the scalar mass $m$ can not be zero to get the nonzero shift. Also, we need to maintain the value of $\lambda$ in such a way that the shift remains small enough, i.e., the absolute value of the correction term inside the square bracket in \eqref{shear_k1} is always less than unity\footnote{In four-dimensional AdS-Schwarzchild blackhole, the momentum of quasinormal mode in the shear channel is a real number. Here we have considered the perturbative correction, i.e., without backreacted background. So its effect should not change the momentum so much that it turns into a complex or imaginary. To constrain this we should not count the correction under square bracket in \eqref{shear_k1} more than unity.}. Next few higher-order pole-skipping points are $\omega_n=-\frac{3}{2}inr_0$ for $n=2,\,3,\dots$ and $k_n$'s are higher orders in $\xi$.
 \begin{eqnarray*}
     k_{2}^{2} &= & 3\sqrt{2}r_0^2 \left[1-\frac{3\lambda\xi\phi(r_0)^p}{4(2\xi+1-\sqrt{2})^{2}}\left(12\xi^{4}+4(21-\sqrt{2})\xi^{3}+(209-74\sqrt{2})\xi^{2}\right.\right.\nn\\
     &&\left.\left.+(134-238\sqrt{2})\xi+136+20\sqrt{2})\right)\right]\label{shear_k2}\\
     k_{3}^{2}&=&  3\sqrt{3}r_0^2\left[1+\frac{(5-\sqrt{3})\lambda\xi\phi(r_0)^p}{66(6\xi-3+2\sqrt{3})^3}\left(-3888\xi^{6}+54432\xi^{5}-(32400-21528\sqrt{3})\xi^{4}\right.\right.\nn\\
     &&\left.\left.+(976140-224964\sqrt{3})\xi^{3}-(1108017-786374\sqrt{3})\xi^{2}+(1134059-427507\sqrt{3})\xi\right.\right. \nn\\
     &&\left.\left.+295381\sqrt{3}-222201\right)\right]\label{shear_k3}
 \end{eqnarray*}
Here we have consciously ignored the background scalar stress-tensor contribution for the numerical analysis to present the effect of interaction clearly. To do this we have assumed $\beta^2\ll\lambda\ll 1$. Anyway, we can consider the background profile and carry out the analysis. We have shown it in the appendix. These same assumptions also continue into the sound mode analysis. In all of these $k$ values, the absolute value of the perturbative correction increases with $\phi(r_0)$ i.e. with $\mathcal{O}_{s,c}$ but the sign of the term is solely decided by the factor $m^2p$. We find that $k_n^2$ can be both greater or less than $3\sqrt{n}r_0^2$ depending on the value of $m^2p$. For example $k_1^2>3r_0^2$ for $\frac{3}{4}\left(1-\sqrt{137}\right)\leq m^2p<-\frac{3}{2}$ and $-\frac{9}{4}\leq m^2\leq -\frac{5}{4}$. However, for other higher mode points, $k_n^2$ is always less than $3\sqrt{n}r_0^2$. It puts no further restriction on scalar mass.  
\begin{figure}[ht]
\centering\includegraphics[width=0.45\textwidth,height=4cm]{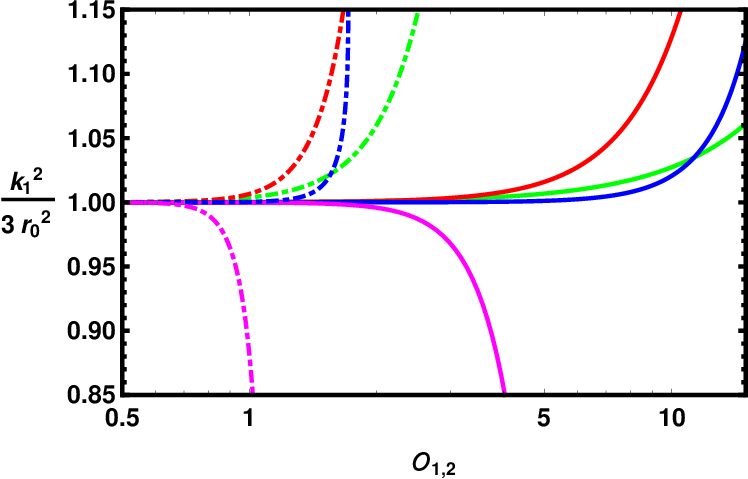}
\includegraphics[width=0.45\textwidth,height=4cm]{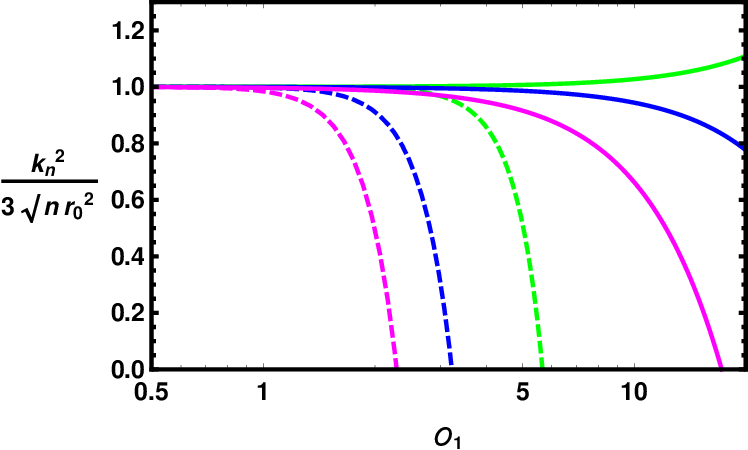}
\caption{{\it Left}: The plot of $\frac{k_1^2}{3 r_0^2}$ vs $\mathcal{O}_2=\frac{\sqrt{|\mathcal{O}_c|}}{|m|}$ (dashed line) and $\mathcal{O}_1=\frac{|\mathcal{O}_s|}{|m|}$ (solid line) where $p=2$ (green color), $p=3$ (red color), $p=4$ (blue color), and $p=5$ (magenta color). {\it Right}: The plot of $\frac{k_n^2}{3\sqrt{n}r_0^2}$ vs $\mathcal{O}_1$ for $n=1$ (green color), $n=2$ (blue color), $n=3$ (magenta color) and for two different powers $p=2$ (solid line) and $p=5$ (dashed line). Here we have taken $\lambda=10^{-5}$ and $m^2=-2.0$}.
\label{shear_plot}
\end{figure}
In \figurename{  \ref{shear_plot}, we have plotted $\frac{k_1^2}{3r_0^2}$. The ratio has been varied with the scalar source $\mathcal{O}_s$ and response $\mathcal{O}_c$ for four different order of interaction $p=2,\,3,\,4\,\&\,5$ with perturbation parameter $\lambda=10^{-5}$ and scalar mass $m^2=-2$. Figure [\ref{shear_plot}] depicts that while the source is off, the ratio is equal to unity. As the source increases from zero, the ratio deviates from unity and increases or decreases according to the power of the Scalar-Gauss-Bonnet interaction term $p$. At the given mass value, for $0<p\leq 4$, the ratio increases with the source, and for $p\geq 5$ the ratio decreases. It has been depicted in the left panel of the figure. In the left plot, we have plotted $k_1^2/(3r_0^2)$ with both $\mathcal{O}_c$ and $\mathcal{O}_s$. This is a very special case from the point of view of the mass of the scalar field $\phi$. When $m^{2}=-2$, both the asymptotic co-efficients ($\mathcal{O}_c$ and $\mathcal{O}_s$) are normalizable modes in standard and alternate quantizations. So, we can take any one of them as the source and examine the effect on the PS points. To visualise this point, we have shown the effect of sources in both quantizations on the momentum values of the shear mode. We have observed that momentum values are very large at smaller source $\mathcal{O}_c$(in alternate quantization)  as compared to source $\mathcal{O}_s$ (in standard quantization). Whereas on the right panel of the same figure, $k_1^2/(3r_0^2)$, $k_2^2/(3\sqrt{2}r_0^2)$ and $k_3^2/(3\sqrt{3}r_0^2)$ have been varied with the scalar source for $p=2$ and $p=5$. $k_1^2/(3r_0^2)$ increases with source $\mathcal{O}_1$ for $p=2$ and decreases for $p=5$ which is consistent with analytical observations as discussed above. On the other hand, $k_2^2/(3\sqrt{2}r_0^2)$ and $k_3^2/(3\sqrt{3}r_0^2)$ decrease with source for both $p=2\,\&\,5$. It is expected to find similar behaviour with $\mathcal{O}_2$.

The boundary Green's functions of these quasi-normal modes have been studied earlier for the boundary theory corresponding to the AdS-Schwarzchild bulk \cite{Huh:2021ppg}. At the hydrodynamic limit, one finds the standard dispersion relation $\omega=\frac{-ik^2}{4\pi T}$ \cite{Blake:2019otz}. On the other hand, in our model, we have found that the $\omega$ value of the pole-skipping point has matched with the first non-hydrodynamic mode of the quasi-normal mode of pure AdS-Schwarzchild case \cite{Huh:2021ppg, Jeong:2021zsv, Arean:2020eus}. Now plotting these two modes -- standard hydrodynamic mode and our pole-skipping points -- we find that the first order pole-skipping point is almost the collision point $(k_{eq},\,\omega_{eq})$ of the hydrodynamic and non-hydrodynamic modes. Therefore, using the first order pole-skipping point, we find the approximate value of the upper bound of diffusion constant \cite{Hartman:2017hhp} and that bound can be written as $\omega_{eq}/(-ik^2_{eq})$. 
Here, in our case, we find the diffusion constant $\mathcal{D}_s$ from pole-skipping as
\begin{equation}\label{shear_diff}
    \mathcal{D}_s=\frac{i\omega_1}{k_1^2}=\frac{1}{2r_0}\left[1 +3\lambda\phi(r_0)^{p}\frac{\xi(2\xi^2-\xi-17)}{2\xi+1}\right]
\end{equation}
Its upper bound is $\mathcal{D}_s(\lambda\neq 0)<\mathcal{D}_s(\lambda=0)$, i.e., the interaction decreases the upper bound for the negative mass regime. The interaction effect gives a higher upper bound for a positive mass regime than the pure AdS-Schwarzchild. For $d+2$ dimensional pure AdS-Schwarzchild black hole,  the diffusion constant is bounded\footnote{For pure Schwarzchild-AdS$_{d+2}$, the shear mode diffusion rate is $\frac{1}{4\pi T}$, where $T=\frac{d+1}{4\pi}r_0$ and $r_0$ is the horizon \cite{Kovtun:2003wp}. So $\mathcal{D}_sT$ is independent of the dimensions of the black hole. The first order pole-skipping point of the shear mode is dimension dependent, $\omega=-\frac{d+1}{2}ir_0$ and $k_1^2=\frac{d(d+1)}{2}r_0^2$. Therefore $\frac{i\omega_1}{k_1^2}=\frac{1}{d\,r_0}=\frac{d+1}{d}\frac{1}{4\pi T}$} as $1\leq 4\pi\mathcal{D}_sT\leq\frac{d+1}{d}$. If the scalar field follows the BF bound and unitarity condition, the scalar mass follows the bound $-2.25<m^2<-1.25$. The term $\mathcal{D}_sT$ in \eqref{shear_diff} can be found in $1\leq4\pi\mathcal{D}_sT\leq \frac{3}{2}$ for all $p\leq 6$ for the mass ranges given in Table \ref{tab_sh}. 
\begin{table}[h]
\centering
\caption{The mass range associated to $p$ to follow the allowed bound of the diffusion coefficient}
\label{tab_sh}
\begin{tabular}{cc}
      Interaction order $p$ ($\phi^p$) & mass range \\
      \hline
   $p = 1$ \hspace{10mm} &  $-2.25 < m^2 < -1.5$\\
   $p = 2$ \hspace{10mm} &  $-2.25 < m^2 < -1.25$\\
   $p = 3$ \hspace{10mm} &  $-2.25 < m^2 < -1.25$\\
   $p = 4$ \hspace{10mm} &  $-2.007 < m^2 < -1.25$\\
   $p = 5$ \hspace{10mm} &  $-1.605 < m^2 < -1.25$\\
   $p = 6$ \hspace{10mm} &  $-1.338 < m^2 < -1.25$\\
   \hline
\end{tabular}
\end{table}
In the allowed mass range, the diffusion constant violates the bounds for $p\geq 7$. 
\begin{figure}[ht]
\begin{minipage}[b]{0.45\linewidth}
\centering
\includegraphics[width=\textwidth,height=4cm]{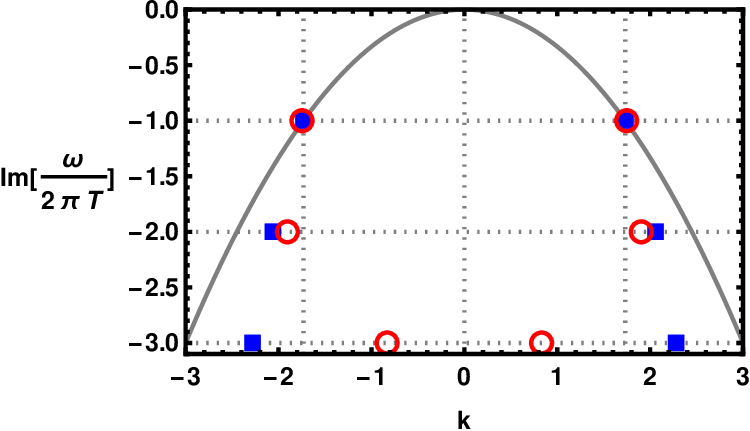}
\end{minipage}
\hspace{0.2cm}
\begin{minipage}[b]{0.45\linewidth}
\centering
\includegraphics[width=\textwidth,height=4cm]{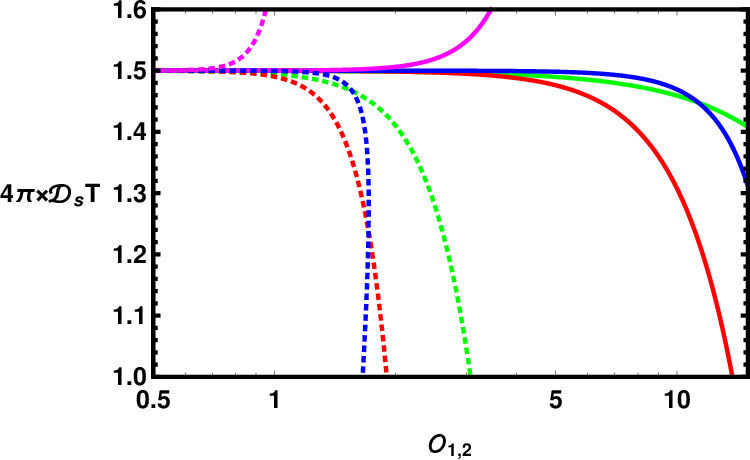}
\end{minipage}
\caption{{\it Left}: The plot of PS points in $\omega-k$ plane for $\lambda=0$ (blue box) and $\lambda=10^{-5}$ (red circle), $\mathcal{O}_1=5.167,\,p=3$ and $m^2=-2$. Three different shapes have been used for three different modes. The solid curve (grey colour) is $\omega=\frac{-ik^2}{4\pi T}$. {\it Right}: Plot of $4\pi\mathcal{D}_sT$ vs $\mathcal{O}_2$ (dashed line) and $\mathcal{O}_1$ (solid line) for $p=2$ (green line), $p=3$ (red line), $p=4$ (blue line) and $p=5$ (magenta line). In all these plots, $m^2=-2.0$ and $\lambda=10^{-5}$}.
\label{shear_plot1}
\end{figure}
In Figure [\ref{shear_plot1}], the left panel shows the plot of the pole-skipping points in the $\omega-k$ plane. Here we have plotted the standard dispersion relation of the boundary theory in a low-frequency regime, $\omega(k)=-i\mathcal{D}_sk^2$ where $\mathcal{D}_s=\frac{1}{4\pi T}$ given in \cite{Blake:2019otz}. When $\lambda=0$ or the perturbative correction is very small, the first pole-skipping point falls on the dispersion curve. As the effect of interaction increases the first pole-skipping point skips the dispersion curve. But, the other pole-skipping points always stay away from the dispersion curve. At the right panel of Figure [\ref{shear_plot1}], we have plotted the diffusion constant obtained in \eqref{shear_diff}. Here the $4\pi\mathcal{D}_sT$ have been varied with the scalar source in standard quantization and in alternate quantization for three different $p$ values. As the source is zero the diffusion constants for all $2\leq p\leq6$ become equal to the upper bound $\frac{3}{8\pi T}$. In the plot, as the source increases from zero, the diffusion constants start falling from the highest bound. For the coupling function $\zeta(\phi)\sim \phi^2$ and $\phi^3$, the diffusion constant decreases monotonically. At a particular value of the sources, $4\pi\mathcal{D}_sT$ has become equal to unity, and for further increase of source value, it has fallen below its lower bound. For $p=4$, the diffusion constant is found to remain very close to its upper bound for a comparatively long range of $\mathcal{O}_{1,2}$. After that, it started decreasing very rapidly and reached below $1$. At these higher values of the source, the $p=4$ curve seems to give two different values of $\mathcal{D}_sT$ at a single value of scalar source $\mathcal{O}_{1,2}$. Again, after a certain value of the source, the diffusion constant violates the lower bound. This issue is important to study in future using full backreacted metric \cite{Buchel:2008vz}. Since our whole calculation is assumed to be in a perturbative regime, we are free to choose any tiny value of $\lambda$ and any small range of the scalar source for the numerical evaluation. Thus the better estimation in our case always makes $1\ll4\pi\mathcal{D}_sT\leq\frac{3}{2}$ for $1<p\leq 6$.    
 

\subsection{Sound Channel}
The longitudinal components of the metric perturbation are called the scalar or sound modes of the perturbation. These are associated with the energy density correlation on the boundary. The corresponding stress-energy tensors in this mode are $T_{vv},\,T_{vx},\,T_{xx}$ and $T_{yy}$ on the boundary field theory. These make the two points correlation functions $G_{vv,vv},\,G_{vv,vx},\,G_{vv,xx}$ and $G_{vv,yy}$ which are induced by the metric perturbations. In holographic gravity theory the required perturbations are $\delta g_{vv},\,\delta g_{vx}$ and $\delta g_{xx}$ with the trace-less perturbation, i.e., $\delta g_{yy}=-\delta g_{xx}$. Like the shear mode, the metric perturbations also combine into a diffeomorphism invariant master variable $\mathcal{Z}_{so}$.
\begin{equation}\label{mastereq}
    \mathcal{Z}_{so}=\frac{1}{r^2}[k^2\d g_{vv}+2\omega\, k \d g_{vx}
    -\frac{k^2}{2}(2f'(r)+rf(r)-\frac{4\omega^2}{k^2})\delta g_{xx}]
\end{equation}
The second-order differential equations of $\delta g_{vv}(r),\,\delta g_{vx}(r)$ and $\delta g_{xx}(r)$ are combined into the master equation.
\begin{equation}\label{so_master_eq}
\mathcal{M}_{so}\mathcal{Z}''_{so}(r)+\mathcal{P}_{so}\mathcal{Z}'_{so}(r)+\mathcal{Q}_{so}\mathcal{Z}_{so}(r)=0
\end{equation}
The coefficients of \eqref{so_master_eq} are linear in $\lambda$ which are given in the appendix. At $\lambda=0$, the master equation reduces to the same for the pure Schwarzchild-AdS$_4$ background. Considering the near horizon structure of $\mathcal{Z}_{so}$ similar to $\mathcal{Z}_{sh}$, we find the pole-skipping points for various orders.

Here we find two types of pole-skipping points from this master equation \eqref{so_master_eq}. The denominator of $\mathcal{P}_{so}$ and $\mathcal{Q}_{so}$ of the above-written master equation contains a common term $3k^2-4\omega^2+k^2f(r)$. At the near horizon regime, it introduces a pole at $3k^2-4\omega^2=0$. Now if we consider $3k^2\neq 4\omega^2$ we get only $\omega_n=-\frac{3}{2}inr_0$ for $n=1,\,2\,\cdots$ at the lower-half plane of complex $\omega$. But when we impose the condition $3k^2=4\omega^2$, we can also find $\omega$ in the upper-half plane of $\omega$, $\omega_n=\frac{3}{2}inr_0$. It will be discussed later. Now we focus on the unequal condition.  

For $3k^2\neq 4\omega^2$, the first order pole-skipping point is found at $\omega_n=-\frac{3}{2}nir_0=-2\pi inT$ and first two $k_n^4$ are given as
\begin{subequations}
    \begin{align}
&\label{sound_k1} k_1^4+9 r_0^4-27\xi\lambda(1+ i) r_0^4 (\xi+2(1+ i)) \phi (r_0)^{p} =0 \\
 &\label{sound_k2} k_2^4+18r_0^4+\frac{9\lambda\xi p r_0^4}{5 \sqrt{2}-2 i} (3\xi ((5 \sqrt{2}-2 i) 3\xi +40 i-64 \sqrt{2})+126 (3 \sqrt{2}-4 i)) \phi (r_0)^{p } = 0 \\
&  k_3^4+27 r_0^4-\frac{6\lambda\xi p r_0^4\phi(r_0)^{p }}{91 \sqrt{3}+63 i} [27(37 \sqrt{3}+3 i)\xi^{3}-189 (61 \sqrt{3}+11 i)\xi^{2}+ 189 (306 \sqrt{3}+31 i)\xi\nn\\
&  -27 (5369 \sqrt{3}+69 i)] = 0 \label{sound_k3}
    \end{align}
\end{subequations}
Higher order $k$ can also be found in the same way. At $\lambda=0$, we get the Schwarzchild-AdS$_4$ values $k_1^4=-9r_0^4,\,k_2^4=-18r_0^4$ and so on. In \eqref{sound_k1}, the imaginary part $3\lambda\left(12+p m^2\right)m^2 p r_0^4\phi(r_0)^p$ is zero for $m^2=-\frac{12}{p}$ which is beyond the BF bound $-\frac{9}{4}<m^2$ for $p\leq 5$. But for $p\geq 6$ we can make $k_1^4$ real at that specific value of $m^2$. A similar behaviour is also expected from the higher order $k$.
\begin{figure}[ht]
\begin{minipage}[b]{0.45\linewidth}
\centering
\includegraphics[width=\textwidth,height=4cm]{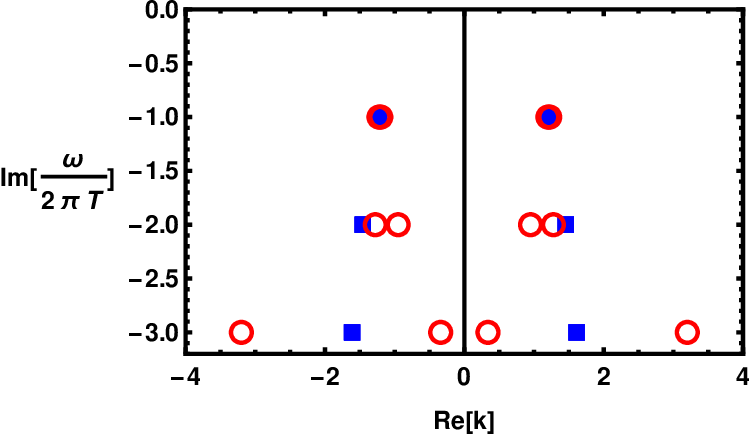}
\end{minipage}
\hspace{0.2cm}
\begin{minipage}[b]{0.45\linewidth}
\centering
\includegraphics[width=\textwidth,height=4cm]{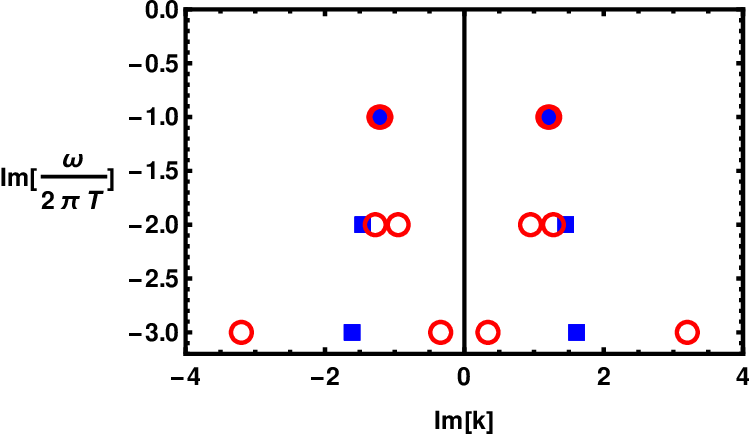}
\end{minipage}
\caption{The plot of real part (right panel) and imaginary part (left panel) of $k$ vs $\text{Im}[\frac{\omega}{2\pi T}]$ for $p=3$, $m^2=-2$, $\mathcal{O}_1=7.234$, $\lambda=0$ (blue rectangle) and $\lambda=10^{-5}$ (red circle).}
\label{sound_plotwk}
\end{figure}
Here we have compared the position of pole-skipping points of $\phi^2$ interaction with the absence of interaction ($\lambda=0$) in Figure \ref{sound_plotwk}. The real and imaginary parts of $k$ have been separately plotted against $\text{Im}[\omega/2\pi T]$. In both cases, the real and imaginary parts are almost equal to each other in each mode. For each part, the values have mirror symmetry with respect to the $\text{Re}[k]=\text{Im}[k]=0$ axes. The shift due to interaction is very hard to identify in $k_1$. For $k_2$ and $k_3$ on the other hand, one observes a measurable amount of shift. It has been depicted in above Figure \ref{sound_plotwk}. Without interaction, in each of these three modes, only four real numbers make four different complex $k$ (having equal absolute values of real and imaginary parts). With interaction, the same happened for $k_1$. But for $k_2$ and $k_3$ eight real numbers make four complex values of $k$.
\begin{figure}[ht]
\begin{center}
\begin{minipage}[b]{0.45\linewidth}
    \includegraphics[width=\textwidth,height=4cm]{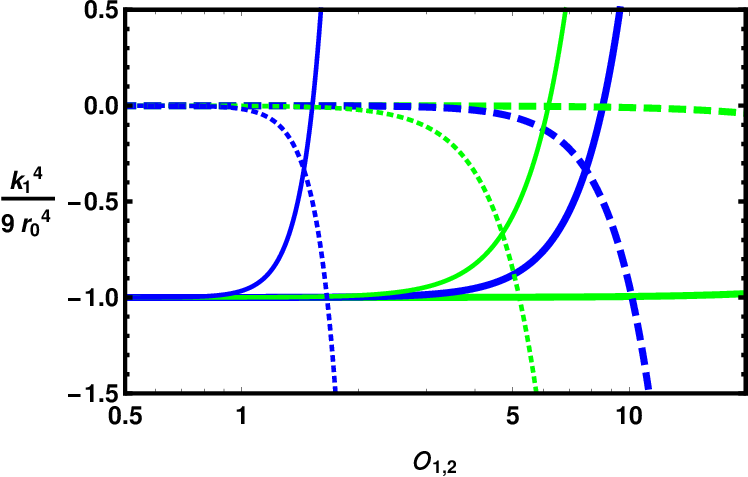}
\end{minipage}
\hspace{0.3cm}
\begin{minipage}[b]{0.45\linewidth}
    \includegraphics[width=\textwidth,height=4cm]{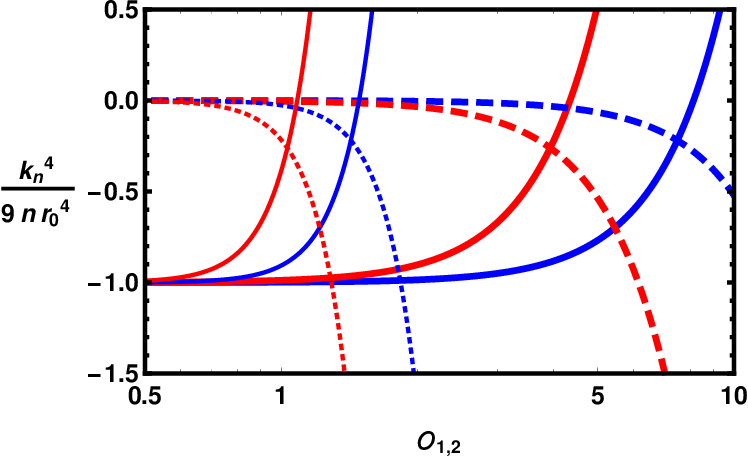}
\end{minipage}
\end{center}
\caption{{\it Left}: The plot of real (solid line) and imaginary (dashed line) parts of $\frac{k_1^4}{9r_0^4}$ against $\mathcal{O}_1$ (thick lines) and $\mathcal{O}_2$ (thin lines) for interaction order of scalar $p=2$ (green colour) and $p=4$ (blue colour) with $\lambda=10^{-5}$. {\it Right}: The plot of real (solid line) and imaginary (dashed line) parts of $\frac{k_n^4}{9r_0^4}$ vs $\mathcal{O}_1$ (thick lines) and $\mathcal{O}_2$ (thin lines) for next two orders of pole-skipping $n=2$ (blue colour) and $n=3$ (red colour); with $\lambda=10^{-5}$ and $p=3$, $m^2=-2$.}
\label{sound_ksource}
\end{figure}
Again we have numerically shown the variation of $k$ with the source $\mathcal{O}_{s}$ in standard quantization and $\mathcal{O}_c$ in alternate quantization of the scalar in \figurename{ \ref{sound_ksource}}. At the left plot of this figure, we have plotted the real and imaginary parts of $k_1^4/(9r_0^4)$ against the source in standard and alternate quantization separately. Here we have evaluated the ratio of our result with the result of pure AdS-Schwarzchild. This ratio has no explicit $r_0$ dependent. It depends on the scalar mass ($m$), interaction order ($p$), and scalar value on the horizon. For $\lambda=10^{-5}$, $m^2=-2$, and for different $p$ the ratio has been evaluated. When the source is off, the imaginary part of the mentioned ratio is zero, whereas the real part is $-1$. Which is consistent with the case without interaction. The imaginary part in $k^4$ is contributed only from the interaction. As we have seen the source is linearly proportional to $\phi(r_0)$, so, $\phi(r_0)$ also goes to zero as the source becomes zero and thus vanishes the correction term in \eqref{sound_k1}. For $p=2\,\&\,4$ we have found the same behaviour as $\mathcal{O}_{1,2}$ go to zero (eventually $\mathcal{O}_s\to 0$). Now if the source is turned on and increased gradually, as long as the source is small enough, both the imaginary and real parts change slowly with $\mathcal{O}_1$. A similar behaviour is found for the variation with $\mathcal{O}_2$, but their variation of $k^4$ is faster. As $p$ increases, the rate of change also increases. The reason is clear from the presence of $\phi(r_0)^p$ factor in the correction terms. During this change the imaginary part of the ratio $k_1^4/9r_0^4$ shifts from $0$ towards $-1$ and the real part changes in the exact opposite direction. Therefore the absolute value of the real (imaginary) part decreases (increases). Thus at some point on $\mathcal{O}_{1,2}$, real and imaginary lines cross each other where their value is exactly equal and lie in between $0$ and $1$. Again after a certain amount of increase in the source, the real part crosses the horizontal axis. At that value $\mathcal{O}_{1,2}$, $k_1^4$ becomes a completely imaginary number. These two cross-over points highly depend on $p$, in the given plot, the $p=4$ plot has made the first cross-over than the $p=2$ plot. As the source value increases further the real (imaginary) values become more and more positive (negative). Since we are interested in the perturbative effect, we will not consider those high values of $k_1^4$. At the right panel of the same figure, we have plotted the ratio $k_n^4/(9nr_0^4)$ for the second ( $n=2$) and third ($n=3$) order pole-skipping. Here interaction order is fixed at $\phi^3$. We have noticed that the behaviour of the real and imaginary parts of the ratio is almost identical to the left panel. We have found that the two cross-overs for each of these two modes of $k$. At these cross-over points, the behaviour of $k_n^2$ is completely identical to before. For the higher order of pole-skipping, $k_n$, the cross-over points come closer to $\mathcal{O}_{1,2}=0$. Therefore the order of interaction and the order of the pole-skipping affect $k$ in the same way. Mainly the location of the cross-over points is almost identically affected by these two parameters. The cross-over points can be found analytically from \eqref{sound_k1}-\eqref{sound_k2}. For example, the real and imaginary parts of $k_1^4$ are $\text{Re}[k_1^4] = -9r_0^4\left(1-\frac{1}{3}\lambda p^2m^4\phi(r_0)^p\right)$ and $\text{Im}[k_1^4] = 3\lambda m^2pr_0^4\left(12+m^2p\right)\phi(r_0)^p$. The first cross-over happens at the values of $\mathcal{O}_{1,2}$ corresponding to $\phi(r_0)=\left(-4\lambda m^2p\right)^{-1/p}$ where the real and imaginary part of $k_1^4$ are equal to each other. The (second) cross-over on the $\mathcal{O}_{1,2}$ axis occurs for $\phi(r_0)=\left(\frac{3}{\lambda p^2m^4}\right)^{1/p}$. Here $k_1^4$ is completely imaginary $9ir_0^4\left(\frac{12}{m^2p}+1\right)$. The first cross-over occurs only if $m^2<0$. For a moment if we assume that $m^2>0$, then there is only the second cross-over where the $k_1^4$ becomes completely imaginary.

\label{sec_chaos}\section{Analysis of chaos}
\subsection{From $vv$ component of linearised Einstein equation}
From the shock wave analysis \cite{Shenker:2014cwa}, it is found that the exponential factor of OTOC can be directly observed from the $\delta E_{00}$ component of the linearized Einstein equation in the ingoing Eddington-Finkelstein co-ordinate. 
In the discussed background \eqref{ef_bh}, the information about OTOC can be obtained from the $vv$ component of the equation \eqref{ein_eq}. Considering the metric perturbation coupled with the $vv$ component of the metric (which are actually the perturbations associated with the sound mode) one can write the $\delta E_{vv}$ at $r=r_0$ as follows.
\begin{equation}
    \delta g_{vv}(r_0)\left(k^{2}-2ir_{0}\omega\right)+k\delta g_{vx}(r_0)\left(2\omega-3ir_{0}\right)=0
\end{equation}
Since it is well-known that at the special points $(\omega_*,\,k_*)$, we have no constraint on the perturbed metric components at $r=r_0$ \cite{Blake:2018leo}. Therefore in the above equation the coefficients of $\delta g_{vv}(r_0)$ and $\delta g_{vx}(r_0)$ have to zero. Thus we have $\omega_*=\frac{3ir_0}{2}=2\pi iT,$ and $k_*^{2}=-3r_{0}^{2}$. This $(\omega_*,k_*)$ is the zeroth order pole-skipping point which is connected to the Lyapunov exponent and butterfly velocity as shown in \eqref{chaos}.
 In our model, we get, $\lambda_{L}=2\pi T$ and $v_{B}=\frac{\sqrt{3}}{2}$, which is the exact result\cite{Blake:2019otz} as in the case of background where the coupling term is not present in the action.

\subsection{From the master equation}
In the last section, where we have discussed the pole-skipping of the sound mode perturbation, we took the condition that $3k^2\neq 4\omega^2$. Because we have seen at the horizon the differential equation \eqref{so_master_eq} encounters a singularity. Here we will discuss that issue. From past works \cite{Blake:2019otz, Wu:2019esr}, we have seen that $3k^2=4\omega^2$ had come with a new set of points $(\omega,\,k)$ in $\text{Im}[\omega]>0$ plane which was actually related to the chaos parameters. In our case,  
we can re-arrange the master equation \eqref{so_master_eq} as 
 \begin{equation}\label{new_master}
     \mathcal{Z}_{so}''(r)+P(r)\mathcal{Z}_{so}'(r)+Q(r)\mathcal{Z}_{so}(r)=0
 \end{equation}
 In this equation, the denominators of both $P(r)$ and $Q(r)$ has a multiplicative factor of $\left(3+f(r)\right)k^2-4\omega^2$ which reduces to $3k^{2}-4\omega^2$ at $r=r_0$. So to get the regular solution of \eqref{new_master} at $r=r_0$, we must impose an extra condition on $\omega$ or $k$. Here we will find it.

First we put $k=\frac{2}{\sqrt{3}}\omega$ in \eqref{new_master} and expand it around $r=r_0$. We find that $P(r)$ and $Q(r)$ possess the first and second order pole at $r=r_0$.
\begin{eqnarray*}
   P(r) & = & \frac{P_{-1}}{(r-r_0)}+\mathcal{O}((r-r_0)^0)\\ P_{-1}&=&-1-\frac{2i\omega}{3r_0}-\frac{144\lambda ir_{0}\omega\zeta'(r_0)}{3r_{0}-2i\omega}\\
Q(r) & = & \frac{Q_{-2}}{(r-r_0)^2}+\mathcal{O}\left((r-r_0)^{-1}\right)\\
    Q_{-2}&=&1+\frac{2i\omega}{3r_0}+\frac{4i\lambda\omega(27r_{0}^{2}+12ir_{0}\omega+4\omega^2)\zeta'(r_0)}{r_{0}(3r_{0}-2i\omega)} 
\end{eqnarray*}
Therefore $r=r_0$ is a regular singular point for the differential equation \eqref{new_master}. Now, suppose $\mathcal{Z}_{so}$ has a series solution near the singular point given as $\mathcal{Z}_{so}=(r-r_0)^{l}\sum_{n\in[0,\mathbb{Z}^+)}\mathcal{Z}_{n}(r-r_0)^{n}$ .
The only condition which makes this solution regular at the horizon is $l=0,1,2,\cdots$. Therefore the first recursion relation coming from \eqref{new_master} is
  \begin{equation}
    l^2+l\left(P_{-1}-1\right)+Q_{-2}=0.
 \end{equation}
This gives two roots (say, $l_1$ and $l_2$) in the following form. 
\begin{eqnarray*}
    l_{1}& = & 1-6\lambda(3r_{0}-2i\omega)\zeta'(r_0)\\
    l_{2}& = & 1+\frac{2i\omega}{3r_0}+6\lambda\frac{(3r_{0}+2i\omega)^{2}}{3r_{0}-2i\omega}\zeta'(r_0)
\end{eqnarray*}
So for arbitrary interaction, the only possible integer roots are $l_1=1$ and $l_2=0$. This gives only two values of $\omega$ as $\pm\frac{3}{2}ir_0$. Therefore we get the same values of the chaos parameters as we have already found in the last subsection.
\par Generally, near-horizon analysis of the $vv$ component of Einstein's equation is adequate to extract the chaotic parameters of the system. In this paper, we have investigated the chaotic property of the system with another approach also: from the singularity analysis of the master equation. We found an exact match between the Lyapunov exponent from both analyses.

\section{Discussions}\label{sec_6}
Here in this paper, we have studied the pole-skipping phenomena in non-extremal gravity theory in the presence of the Scalar-Gauss-Bonnet interaction. We have considered a four-dimensional Schwarzchild-AdS black hole solution as the holographic bulk theory. On the boundary, we have a finite temperature conformal theory. The interaction is sourced by an operator of dimension $\Delta$ of the boundary theory, which is dual to the scalar field $\phi$ in the bulk. In the Einstein action, the interaction term is added perturbatively \eqref{e_action}. In the perturbative approximation, this external scalar source does not affect the original bulk solution but makes a nontrivial contribution in the linearised field equations \eqref{ein_eq}. We have found that $k$ of the pole-skipping points ($\omega,\,k$) corresponding to the scalar field and metric perturbation have been affected by the external scalar source $\mathcal{O}_s$ in both quantizations, whereas $\omega$ remains unchanged. 
\par Unlike the unperturbed model, the minimally coupled scalar $\phi$ has contained both real and imaginary $k$ in the pole-skipping points. As the source is increased, the points of the imaginary $k$ plane have moved into the real $k$ plane. We have presented these facts pictorially in Figure \ref{scal_plot}. In Schwarzchild-AdS$_4$ without external effect \cite{Blake:2019otz}, $k$ is always real in the shear mode. Here we have found that the shear mode $k$ can have both real and imaginary values depending on the effect of the scalar source. We have analytically found the effect of the interaction on the poles located at $\omega_n=-2in\pi T$ and corresponding $k\sim T$ which are given in \eqref{shear_k1} and higher order momentum values of shear mode.  The first order pole-point $k_1^2$ is always greater than $3r_0^2$ for $\zeta=\phi,\,\phi^2,\,\phi^3\,\&\,\phi^4$ and has decreased for other higher powers of $\phi$. However, for the second and other higher orders of pole-skipping, $k^2$ has always decreased with the increasing source for all positive integer powers of $\phi$ in $\zeta(\phi)$. These have been shown in Figure \ref{shear_plot}. Here, the increase (or decrease) of real $k$ implies a slow (or fast) rate of momentum transportation in shear mode and the imaginary $k$ means the exponential decay of the momentum density. As a result, when positive $k_1^2$ has increased with the increasing source $\mathcal{O}_s$, the mobility of the corresponding modes has decreased. Thus the decreasing mobility has decreased the value of diffusion coefficient $\mathcal{D}_s$. In \figurename{ \ref{shear_plot1}}, we have presented this consistent behaviour of diffusion coefficient. At $\mathcal{O}_s\to 0$, $k_1^2$ is at a minimum value, and therefore, momentum flow is maximum which has given the maximum value of $\mathcal{D}_s$. So, due to the effect of the external source, the flow of momentum in shear mode has decreased for $p\leq 4$, otherwise, it has increased. The 1st-order pole-skipping point lies in the dispersion curve. 
\par In the sound mode, the first three pole-skipping points have been derived from the master equation as $\omega_n=-2\pi inT$ and corresponding $k_n\sim T$ is given in \eqref{sound_k1} and \eqref{sound_k2}. In the non-perturbative case where either $\lambda\to 0$ or $\mathcal{O}_s\to 0$, our results have reduced into the pole-skipping points of pure Schwarzchild-AdS$_4$ background \cite{Blake:2019otz}, i.e, $k_n^4=-9nr_0^4$. It gives a complex value (of equal real and imaginary parts) of $k$. As the source is turned on, we have found that an imaginary part has been added with the negative real part of $k^4$. It means the real and imaginary parts of $k$ are no longer equal. We have shown all of these in Figure \ref{sound_ksource}. From the OTOC calculation in the last section, we have found the Lyapunov exponent $\lambda_{L}=i\omega=2\pi T$ and the butterfly velocity $v_b=\frac{\sqrt{3}}{2}$ where $\omega_*=2i\pi T$ and $k_*=\pm\frac{4}{\sqrt{3}}i\pi T$. These results have been further verified with a different approach by analyzing the power series solution of the sound mode master equation near the horizon. Therefore ($\omega_*,\,k_*$) is considered as the lowest order pole-skipping point in sound mode instead of ($\omega_1,\,k_1$). So the pole-skipping points of sound mode are ($\omega_*,\,k_*$), ($\omega_1,\,k_1$), ($\omega_2,\,k_2$), ($\omega_3,\,k_3$) and so on. The pole-skipping points $(\omega,\,k)$ describe the flow of energy density. Here $k$ has both the real and imaginary parts. It signifies that the real part is associated with the flow of the energy density in longitudinal mode whereas the imaginary part of $k$ is related to the exponential decay of the energy density. Therefore with the effect of interaction, when the energy density diffusion has increased, the exponential decay has decreased and vice-versa. It would be interesting to study these flows and decays quantitatively. It would be very interesting to calculate the bulk-boundary Green's function with the set-up of this paper and calculate the pole-skipping points from Green's function. 
\par In this work, we have found some non-trivial effects of the interaction on the sound mode and shear mode. We have not found any effect on the chaotic behaviour. The reason is mainly the perturbative approach to the interaction term. If one considers the backreaction of the interaction, the Lyapunov exponent and the butterfly velocity are expected to be affected by the interaction. With backreaction, one can expect $k_*$ and $k_1$ to be equal in the sound mode.

\vskip 3mm
\noindent
{\bf Acknowledgements}
\vskip 3mm
\noindent

We would like to acknowledge Debaprasad Maity for his useful suggestions. BB would like to acknowledge the MHRD, Govt. of India for providing the necessary funding and fellowship to pursue research work.

\appendix

\section{Coefficient of Master Equation: Shear Channel}\label{apnd_shear}
Three coefficients of the master equation can be written in the linear order of the perturbation parameter $\lambda$
\begin{eqnarray}
\mathcal{M}_{sh}(r) & = & \mathcal{M}_{sh}^{(0)}+\lambda\mathcal{M}_{sh}^{(1)}+\mathcal{O}(\lambda^2)\nn\\
\mathcal{P}_{sh}(r) & = & \mathcal{P}_{sh}^{(0)}+\beta^{2}\mathcal{P}_{sh}^{(\phi)}+\lambda\mathcal{P}_{sh}^{(1)}+\mathcal{O}(\lambda^2)\nn\\
\mathcal{Q}_{sh}(r) & = & \mathcal{Q}_{sh}^{(0)}+\beta^{2}\mathcal{Q}_{sh}^{(\phi)}+\lambda\mathcal{Q}_{sh}^{(1)}+\mathcal{O}(\lambda^2)\nn
\end{eqnarray}
We have found the above functions as follows.
\begin{eqnarray}
\mathcal{M}_{sh}^{(0)} & = & r^2 f(r)\\
\mathcal{P}_{sh}^{(0)} & = & \frac{\omega  f(r) \left(5 r \omega +2 i k^2\right)-8 k^2 r f(r)^2+\omega ^2 (3 r-2 i \omega )}{\omega ^2-k^2 f(r)}\\
\mathcal{Q}_{sh}^{(0)} & = & \frac{-10 k^2 r^2 f(r)^2+f(r) \left(k^4+9 i k^2 r \omega +4 r^2 \omega ^2\right)+\omega  \left(k^2 (-\omega -3 i r)+6 r \omega  (r-i \omega )\right)}{r^2 \left(\omega ^2-k^2 f(r)\right)}\nn\\
\end{eqnarray}
and
\begin{eqnarray}
\mathcal{M}_{sh}^{(1)} & = & 0\\
\mathcal{P}_{sh}^{(1)} & = & \frac{r^2f(r)}{\left(\omega ^2-k^2 f(r)\right)^2}\left[r \zeta ''(r) \left(\omega ^2-k^2 f(r)\right) \left(f(r) \left(2 k^2 f(r)+\omega ^2\right)-3 \omega ^2\right)+\zeta '(r) \left(f(r) \right.\right.\nn\\
&& \left.\left. \left.(k^2 f(r) \left(4 k^2 f(r)-6 k^2 -11 \omega ^2\right)+24 k^2 \omega ^2-2 \omega ^4\right)-9 k^2 \omega ^2\right)-\omega\mathcal{F}\right]\\
\mathcal{Q}_{sh}^{(1)} & = & \frac{1}{r\left(\omega^2-k^2 f(r)\right)^2}\left[r \zeta ''(r) \left(\omega ^2-k^2 f(r)\right) \left(f(r) \left(\omega ^2 \left(4 k^2-i r \omega \right)-2 k^2 f(r) \left(-3 r^2 f(r)\right.\right.\right.\right.\nn\\
&&+k^2+3 r^2+i r \omega\left. +\omega ^3 (-2 \omega +3 i r)\right)+\zeta '(r) \left(f(r) \left(f(r) \left(-k^2 f(r) \left(-14 k^2 r^2 f(r)+3 k^4\right.\right.\right.\right.\nn\\
&&\left.\left.   +4 k^2 r (6 r+i \omega )+34 r^2 \omega ^2\right)+3 k^6+6 k^4 \left(3 r^2+i r \omega +\omega ^2\right)+k^2 r \omega ^2 (72 r+11 i \omega )  \right.\nn\\
&& \left.\left.\left.  +2 r^2 \omega ^4\right)+\omega ^2\left(-6 k^4-3 k^2 \left(18 r^2+8 i r \omega +\omega ^2\right)+2 r \omega (-6 r+i \omega )\right)\right)+3 \omega ^3 \left(6 r^2 \omega \right.\right.\nn\\
&&\left.\left.\left.+k^2 (\omega +3 i r)\right)\right)+i r \left(2 \omega ^2-f(r) \left(k^2-2 i r \omega \right)\right) \mathcal{F}\right]
\end{eqnarray}
where,
\begin{eqnarray*}
     \mathcal{F} & = & 6 k^2 r^2 \omega (f(r)-1)\left[r (f(r)-3) f(r) \zeta ''(r)-3 ((f(r)-2) f(r)+3) \zeta '(r)\right]^2/\left[r \zeta ''(r) \left(f(r) \right.\right.\nn\\
     &&\left.\left.\left(f(r) \left(k^2-3 i r \omega \right)  -3 k^2+3 i r \omega \right)-18 i r \omega \right)-3 \zeta '(r) \left((f(r)-2) f(r) \left(k^2-i r \omega \right)\nn\right.\right.\\
     &&\left.\left.+3 \left(k^2+3 i r \omega \right)\right)  +i r^3 \omega  (f(r)-3) f(r) \zeta'''(r)\right]
\end{eqnarray*}

\begin{eqnarray*}
     \mathcal{P}_{sh}^{(\phi)}& = &\frac{1}{(\omega^{2}-k^{2}f(r))^{3}\Xi}\left(-3 k^2 r^5 \omega  (f(r)-1) f(r)^3 \phi '(r)^3 \left(i \omega  f(r) \left(4 \omega ^2 f(r)^2 \left(\omega ^2\nn\right.\right.\right.\right.\\&&\left.\left.\left.\left.-k^2 f(r)\right) \zeta ''(r) \phi ''(r)^2 r^4+4 \omega  f(r) \left(3 r \omega ^3+\left(3 r \omega  (\omega-4 i r)-k^2 (3 r+i \omega )\right) f(r) \omega\nn\right.\right.\right.\right.\\&&\left.\left.\left.\left. +i \left(k^4+3 i r \omega  k^2+4 r^2 \omega ^2\right) f(r)^2\right) \phi '(r) \zeta ''(r) \phi ''(r) r^2+\phi '(r)^2 \left(\left(9 r^2 \omega ^4+3 r \left(6 r \omega
   ^2\nn\nn\right.\right.\right.\right.\right.\right.\\&&\left.\left.\left.\left.\left.\left.-k^2 (3 r+2 i \omega )\right) f(r) \omega ^2+\left((6 i r-\omega ) k^4-6 r (5 r+i \omega ) \omega  k^2+3 r^2 \omega ^2 (3 \omega -20 i r)\right) f(r)^2 \omega\right.\right.\right.\right.\right.\\&&\left.\left.\left.\left.\left. +\left(k^6+6 i r \omega  k^4-5 r^2 \omega ^2 k^2+36i r^3 \omega ^3\right) f(r)^3\right) \zeta ''(r)-4 i r^4 \omega ^3 (f(r)-3) f(r)^2 \zeta ^{(3)}(r)\right)\right) r^2\nn\right.\right.\\&&\left.\left.+\zeta '(r) \left(4 \omega ^2 f(r)^2 \left((3 i r+\omega ) \omega ^3+\left(k^4+i r \omega k^2\right) f(r)^2+i \left(k^2 (3 r+2 i \omega ) \omega -7 r \omega ^3\right) f(r)\right)\nn\right.\right.\right.\\&&\left.\left.\left. \phi ''(r)^2 r^4+4 \omega  f(r) \left(3 r (3 i r+\omega ) \omega ^4+\left(i \left(9 r^2+3 i \omega  r-\omega ^2\right) k^2+3 r \omega  \left(12 r^2-4 i \omega  r\nn\right.\right.\right.\right.\right.\right.\\&&\left.\left.\left.\left.\left.\left. +\omega ^2\right)\right) f(r) \omega ^2+\left((6 r+2 i \omega ) k^4 +r (12 i r-13 \omega ) \omega  k^2-3 r^2 (8 r+7 i \omega ) \omega ^2\right) f(r)^2 \omega +\left(-ik^6\nn\right.\right.\right.\right.\right.\\&&\left.\left.\left.\left.\left.+4 r \omega  k^4+3 i r^2 \omega ^2 k^2+12 r^3 \omega ^3\right) f(r)^3\right) \phi '(r) \phi ''(r) r^2+\left(9 r^2 (3 i r+\omega ) \omega ^5+3 r \left(i \left(9 r^2\nn\right.\right.\right.\right.\right.\right.\\&&\left.\left.\left.\left.\left.\left.-2 \omega ^2\right) k^2+3 r \omega  \left(24 r^2-i \omega  r+2 \omega ^2\right)\right) f(r) \omega ^3+\left(\left(27 r^2+9 i \omega  r-\omega ^2\right) k^4+3 i r \omega  \left(9 r^2\nn\right.\right.\right.\right.\right.\right.\\&&\left.\left.\left.\left.\left.\left.+20 i \omega  r-2 \omega ^2\right) k^2+9 r^2 \omega ^2 \left(-4 r^2-11 i\omega  r+\omega ^2\right)\right) f(r)^2 \omega ^2+\left((2 \omega -9 i r) k^6\nn\right.\right.\right.\right.\right.\\&&\left.\left.\left.\left.\left.+r (42 r+19 i \omega ) \omega  k^4+3 r^2 (23 i r-20 \omega ) \omega ^2 k^2-3 r^3 (32 r+21 i \omega ) \omega ^3\right) f(r)^3 \omega\nn\right.\right.\right.\right.\\&&\left.\left.\left.\left.
   -\left(k^8+7 i r \omega  k^6-15 r^2 \omega ^2 k^4+3 i r^3 \omega ^3 k^2-84 r^4 \omega ^4\right) f(r)^4\right) \phi '(r)^2\right)\right)\right)\\
   \mathcal{Q}_{sh}^{(\phi)}&=&\frac{r^{3}f(r)\phi'(r)^2}{(\omega^{2}-k^{2}f(r))^{3}\Xi}\left(3 k^2 r^2 \omega  (f(r)-1) \left(\left(k^2-2 i r \omega \right) f(r)-2 \omega ^2\right) \phi '(r) \left(4 \omega ^2 f(r)^2 \left(\omega ^2\nn\right.\right.\right.\\&&\left.\left.\left.-k^2 f(r)\right) \zeta ''(r) \phi ''(r)^2 r^4+4 \omega  f(r) \left(3
   r \omega ^3+\left(3 r \omega  (\omega -4 i r)-k^2 (3 r+i \omega )\right) f(r) \omega +i \left(k^4+3 i r \omega  k^2\nn\right.\right.\right.\right.\\&&\left.\left.\left.\left.+4 r^2 \omega ^2\right) f(r)^2\right) \phi '(r) \zeta ''(r) \phi ''(r) r^2+\phi '(r)^2
   \left(\left(9 r^2 \omega ^4+3 r \left(6 r \omega ^2-k^2 (3 r+2 i \omega )\right) f(r) \omega ^2\nn\right.\right.\right.\right.\\&&\left.\left.\left.\left.+\left((6 i r-\omega ) k^4-6 r (5 r+i \omega ) \omega  k^2+3 r^2 \omega ^2 (3 \omega -20 i r)\right) f(r)^2
   \omega +\left(k^6+6 i r \omega  k^4-5 r^2 \omega ^2 k^2\nn\right.\right.\right.\right.\right.\\&&\left.\left.\left.\left.\left.+36 i r^3 \omega ^3\right) f(r)^3\right) \zeta ''(r)-4 i r^4 \omega ^3 (f(r)-3) f(r)^2 \zeta ^{(3)}(r)\right)\right) f(r)^2+\zeta '(r) \left(8 \omega
   ^3 (f(r)-3) f(r)^3 \left(\omega ^2\nn\right.\right.\right.\\&&\left.\left.\left.-k^2 f(r)\right)^3 \phi ''(r)^3 r^6+12 i \omega ^2 f(r)^2 \left(9 i r \omega ^7+\left((21 i r-4 \omega ) k^4+\omega  \left(6 r^2-6 i \omega  r+\omega ^2\right) k^2\nn\right.\right.\right.\right.\\&&\left.\left.\left.\left.-3 i r
   \omega ^4\right) f(r)^2 \omega ^3+2 k^4 r \left(2 i k^2-r \omega \right) f(r)^5 \omega +k^2 \left((5 \omega -6 i r) k^4-2 \omega  \left(-3 r^2-11 i \omega  r+\omega ^2\right) k^2\nn\right.\right.\right.\right.\\&&\left.\left.\left.\left.-20 r^2 \omega ^3\right)
   f(r)^3 \omega +\left(-2 k^8+\omega  (\omega -10 i r) k^6-r (4 r+7 i \omega ) \omega ^2 k^4+14 r^2 \omega ^4 k^2\right) f(r)^4+\left(6 i r \omega ^7\nn\right.\right.\right.\right.\\&&\left.\left.\left.\left.+k^2 (\omega -30 i r) \omega ^5\right) f(r)\right) \phi
   '(r) \phi ''(r)^2 r^4-6 \omega  f(r) \left(27 r^2 \omega ^8+3 r \left(15 r \omega ^2-k^2 (33 r+2 i \omega )\right) f(r) \omega ^6\nn\right.\right.\right.\\&&\left.\left.\left.+\left(\left(45 r^2+30 i \omega  r+\omega ^2\right) k^4+3 i r \omega 
   \left(36 r^2+27 i \omega  r-4 \omega ^2\right) k^2+9 r^2 \omega ^4\right) f(r)^2 \omega ^4-\left(\left(9 r^2+18 i \omega  r\nn\right.\right.\right.\right.\right.\\&&\left.\left.\left.\left.\left.+\omega ^2\right) k^6+3 \omega  \left(36 i r^3-37 \omega  r^2-4 i \omega ^2
   r+\omega ^3\right) k^4+3 r \omega ^2 \left(48 r^3+52 i \omega  r^2+3 \omega ^2 r+2 i \omega ^3\right) k^2\nn\right.\right.\right.\right.\\&&\left.\left.\left.\left.+9 r^2 \omega ^6\right) f(r)^3 \omega ^2+k^2 \left((6 i r-\omega ) k^6+\omega  \left(-63 r^2-44 i
   \omega  r+7 \omega ^2\right) k^4+r \omega ^2 \left(108 i r^2+103 \omega  r\nn\right.\right.\right.\right.\right.\\&&\left.\left.\left.\left.\left.+30 i \omega ^2\right) k^2+3 r^2 \omega ^3 \left(80 r^2+60 i \omega  r-9 \omega ^2\right)\right) f(r)^4 \omega +k^2 \left(k^8-2 i
   r \omega  k^6+19 r^2 \omega ^2 k^4+36 i r^3 \omega ^3 k^2\nn\right.\right.\right.\right.\\&&\left.\left.\left.\left.+48 r^4 \omega ^4\right) f(r)^6+k^2 \left(k^8+5 (4 i r-\omega ) \omega  k^6-r (19 r+10 i \omega ) \omega ^2 k^4+r^2 (-36 i r-43 \omega ) \omega ^3
   k^2\nn\right.\right.\right.\right.\\&&\left.\left.\left.\left.-12 r^3 (12 r+11 i \omega ) \omega ^4\right) f(r)^5\right) \phi '(r)^2 \phi ''(r) r^2+\left(-81 r^3 \omega ^9+27 r^2 \left(k^2 (12 r+i \omega )-8 r \omega ^2\right) f(r) \omega ^7\nn\right.\right.\right.\\&&\left.\left.\left.-9 r \left(\left(9
   r^2+18 i \omega  r+\omega ^2\right) k^4+9 i r \omega  \left(14 r^2+6 i \omega  r-\omega ^2\right) k^2+18 r^2 \omega ^4\right) f(r)^2 \omega ^5+3 k^2 \left(i \left(9 r^2+\omega ^2\right) k^4\nn\right.\right.\right.\right.\\&&\left.\left.\left.\left.+6 r \left(45 i
   r^3-30 \omega  r^2-6 i \omega ^2 r+\omega ^3\right) k^2+27 r^2 \omega  \left(16 r^3+16 i \omega  r^2+2 \omega ^2 r+i \omega ^3\right)\right) f(r)^3 \omega ^4-\left((9 r\nn\right.\right.\right.\right.\\&&\left.\left.\left.\left.+6 i \omega ) k^8+\left(-378 r^3-243
   i \omega  r^2+18 \omega ^2 r+5 i \omega ^3\right) k^6+27 r \omega  \left(28 i r^3+18 \omega  r^2+6 i \omega ^2 r-\omega ^3\right) k^4\nn\right.\right.\right.\right.\\&&\left.\left.\left.\left.+9 r^2 \omega ^2 \left(168 r^3+20 i \omega  r^2-18 \omega ^2 r-3 i
   \omega ^3\right) k^2-27 r^3 \omega ^6\right) f(r)^4 \omega ^3+3 k^2 \left(i k^8+\left(-48 i r^2+22 \omega  r\nn\right.\right.\right.\right.\right.\\&&\left.\left.\left.\left.\left.+4 i \omega ^2\right) k^6+r \omega  \left(42 r^2+119 i \omega  r-30 \omega ^2\right) k^4+12 i
   r^2 \omega ^2 \left(2 r^2+15 i \omega  r-6 \omega ^2\right) k^2-6 r^3 \omega ^3 \left(20 r^2\nn\right.\right.\right.\right.\right.\\&&\left.\left.\left.\left.\left.+48 i \omega  r-9 \omega ^2\right)\right) f(r)^5 \omega ^2+3 k^2 \left(-(10 r+3 i \omega ) k^8+r \omega  (17
   \omega -32 i r) k^6+r^2 (18 r+7 i \omega ) \omega ^2 k^4\nn\right.\right.\right.\right.\\&&\left.\left.\left.\left.+9 r^3 \omega ^3 (4 i r+13 \omega ) k^2+6 r^4 (60 r+49 i \omega ) \omega ^4\right) f(r)^6 \omega +2 i k^2 \left(k^{10}+3 i r \omega  k^8+12 r^2
   \omega ^2 k^6\nn\right.\right.\right.\right.\\&&\left.\left.\left.\left.+63 i r^3 \omega ^3 k^4-117 r^4 \omega ^4 k^2+252 i r^5 \omega ^5\right) f(r)^7\right) \phi '(r)^3\right)\right).\\
   \Xi&=&(3 r\omega -(i k^2 + 3 r \omega)f(r)) \phi'(r) + 2 r^2\omega f(r)\phi''(r).
\end{eqnarray*}

Here the $\zeta$ function takes its appropriate form as mentioned earlier.
After the near-horizon analysis of the master equation, we found the first pole-skipping point $(\omega_1,k_1)$ at $\omega_1=-\frac{3}{2}ir_0$ and 
\begin{eqnarray}
& k_{1}^{2}-3r_{0}^2-\frac{9 \beta^{2}  \xi r_{0}^2 \phi (r_0)^2 \left(r_{0}^2 (3 \xi +4 p)-2 k^2 p\right)}{p \left(2 k^2 p+3 r_{0}^2 (3 \xi +4 p)\right)}+\frac{3 \lambda  \xi  \phi (r_0)^p}{\left(2 k^2 p+3 r_{0}^2 (3 \xi +4 p)\right)^2} \left[-4 k^6 p^2+12 k^4 p r_{0}^2 ((\xi -12) p-3 \xi )\right.\nn\\&\left.+9 k^2 r_{0}^4 \left(-9 \xi ^2+16 (3 \xi -10)
   p^2+12 (\xi -10) \xi  p\right)+27 \xi r_{0}^6 (3 \xi +4 p)^2\right]=0
\end{eqnarray}
Similarly, one can find the higher-order pole-skipping points by performing the near-horizon expansion of the master equation. Here, by considering the negligible value of the $\beta$ parameter, one can get the master equation with interaction only. However, in the shear mode case, to get the previous result \eqref{shear_k1}, one needs to ignore the $\beta^2$ term from the starting point.

\section{Coefficient of Master Equation: Sound Channel}\label{apnd_sound}
Three coefficients of the master equation can be written in the linear order of the perturbation parameter $\lambda$
\begin{eqnarray*}
\mathcal{M}_{so}(r) & = & \mathcal{M}_{so}^{(0)}+\lambda\mathcal{M}_{so}^{(1)}+\mathcal{O}(\lambda^2)\nn\\
\mathcal{P}_{so}(r) & = & \mathcal{P}_{so}^{(0)}+\beta^{2}\mathcal{P}_{so}^{\phi}+\lambda\mathcal{P}_{so}^{(1)}+\mathcal{O}(\lambda^2)\nn\\
\mathcal{Q}_{so}(r) & = & \mathcal{Q}_{so}^{(0)}+\beta^{2}\mathcal{Q}_{so}^{\phi}+\lambda\mathcal{Q}_{so}^{(1)}+\mathcal{O}(\lambda^2)
\end{eqnarray*}
where,
\begin{eqnarray}
\mathcal{M}_{so}^{(0)} & = & r^4 f(r)\\
\mathcal{P}_{so}^{(0)} & = & \frac{r^2 \left(f(r) \left(11 k^2 r f(r)+2 k^2 (6 r-i \omega )-20 r \omega ^2\right)+\left(3 k^2-4 \omega ^2\right) (3 r-2 i \omega )\right)}{k^2 f(r)+3 k^2-4 \omega ^2}\\
\mathcal{Q}_{so}^{(0)} & = & \frac{1}{k^2 f(r)+3 k^2-4 \omega ^2}\left(-f(r) \left(-25 k^2 r^2 f(r)+k^4+12 k^2 r (r+i \omega )+16 r^2 \omega ^2\right)-3 k^4\right.\nn\\
&&\left.\hspace{1cm}+k^2 (9 r+2 i \omega ) (3 r-2 i \omega )
-24 r \omega ^2 (r-i \omega )\right)
\end{eqnarray}

\begin{eqnarray}
    \mathcal{P}_{so}^{\phi} & = & \left[2 k^2 r f(r)^2 \left(r \phi '(r)^2 \left(k^2 r f(r)^2 \left(k^4-4 k^2 r (3 r+i \omega )+16 r^2 \omega ^2\right)-2 f(r) \left(k^6 (6 r+i \omega ) \right.\right.\right.\right.\nn\\
    &&\left.\left.\left.\left. -2 k^4 r \left(9 r^2+3 i r \omega +2 \omega ^2\right)+8 k^2 r^2 \omega ^2 (6 r+i \omega )-32 r^3 \omega ^4\right)+k^2 \left(3 k^2-4 \omega ^2\right) \left(4 r \omega ^2 \right.\right.\right.\right.\nn\\
    &&\left.\left.\left.\left. -k^2 (3 r+2 i \omega )\right)\right)-4 m^2 \phi (r)^2 \left(k^2 r f(r) \left(-4 r \omega ^2+k^2 (3 r+i \omega )\right)+3 k^6-k^4 \left(9 r^2 \right.\right.\right.\right.\nn\\
    &&\left.\left.\left.\left. +3 i r \omega +2 \omega ^2\right)+4 k^2 r \omega ^2 (6 r+i \omega )-16 r^2 \omega ^4\right)\right)\right]/\left\{\left(k^2+2 i r \omega \right) \left(k^2 f(r)+3 k^2-4 \omega ^2\right)^3\right\}\nn\\
    &&\\
    \mathcal{Q}_{so}^{\phi} & = & -\left[r \phi '(r)^2 \left(2 k^4 r^3 f(r)^4 \left(-9 k^4+6 k^2 r (20 r+7 i \omega )-160 r^2 \omega ^2\right)+k^2 r f(r)^3 \left(k^8 \right.\right.\right.\nn\\
    &&\left.\left.\left. +2 k^6 r (93 r+17 i \omega )-4 k^4 r^2 \left(108 r^2+39 i r \omega +20 \omega ^2\right)+48 k^2 r^3 \omega ^2 (20 r+3 i \omega )-512 r^4 \omega ^4\right) \right.\right.\nn\\
    &&\left.\left. +k^2 f(r)^2 \left(-k^8 (9 r+2 i \omega )+6 k^6 r \left(27 r^2+9 i r \omega +2 \omega ^2\right)+4 i k^4 r^2 \omega  \left(63 r^2+144 i r \omega -32 \omega ^2\right) \right.\right.\right.\nn\\
    &&\left.\left.\left. +96 k^2 r^3 \omega ^3 (4 \omega -7 i r)+448 i r^4 \omega ^5\right)+f(r) \left(3 k^2-4 \omega ^2\right) \left(-k^8 (15 r+4 i \omega )+2 k^6 r \left(9 r^2 \right.\right.\right.\right.\nn\\
    &&\left.\left.\left.\left. -3 i r \omega +10 \omega ^2\right)+12 i k^4 r^2 \omega  \left(3 r^2+10 i r \omega -2 \omega ^2\right)+96 k^2 r^3 \omega ^3 (\omega -i r)+64 i r^4 \omega ^5\right) \right.\right.\nn\\
    &&\left.\left. -k^4 \left(3 k^2-4 \omega ^2\right)^2 (3 r+2 i \omega ) \left(k^2+2 i r \omega \right)\right)+2 m^2 \phi (r)^2 \left(k^4 r^2 f(r)^3 \left(k^4+6 k^2 r (20 r+7 i \omega ) \right.\right.\right.\nn\\
    &&\left.\left.\left. -160 r^2 \omega ^2\right)+k^2 r f(r)^2 \left(k^6 (105 r-2 i \omega )-2 k^4 r \left(108 r^2+39 i r \omega +32 \omega ^2\right)+24 k^2 r^2 \omega ^2 (20 r \right.\right.\right.\nn\\
    &&\left.\left.\left. +3 i \omega )-256 r^3 \omega ^4\right)+f(r) \left(-6 k^{10}+k^8 \left(27 r^2-24 i r \omega +4 \omega ^2\right)+2 i k^6 r \omega  \left(63 r^2+84 i r \omega +8 \omega ^2\right) \right.\right.\right.\nn\\
    &&\left.\left.\left. +16 k^4 r^2 \omega ^3 (8 \omega -21 i r)+224 i k^2 r^3 \omega ^5\right)-\left(3 k^2-4 \omega ^2\right) \left(6 k^8-k^6 \left(9 r^2+6 i r \omega +4 \omega ^2\right) \right.\right.\right.\nn\\
    &&\left.\left.\left. +2 k^4 r \omega  \left(-9 i r^2+30 r \omega +4 i \omega ^2\right)+48 i k^2 r^2 \omega ^3 (r+i \omega )-32 i r^3 \omega ^5\right)\right)\right]/\nn\\
    && \left\{2 r^2 \left(k^2+2 i r \omega \right) \left(k^2 f(r)+3 k^2-4 \omega ^2\right)^3\right\}
\end{eqnarray}
}
\begin{eqnarray}
\mathcal{M}_{so}^{(1)} & = & 0\\
\mathcal{P}_{so}^{(1)} & = & \frac{r^2 f(r)}{\left(2 i r \omega + k^2\right) \left(k^2 f(r)+3 k^2-4 \omega ^2\right)^3} \left[r^3 \zeta ''(r) \left\{f(r) \left(k^2 f(r) \left(k^2 f(r)^2 \left(9 k^4\right.\right.\right.\right.\right.\nn\\
&&\left. +2 k^2 r (-24 r+25 i \omega )+64 r^2 \omega ^2\right)+2 f(r) \left(-27 k^6+6 k^4 \left(24 r^2-9 i r \omega+11 \omega ^2\right) \right.\nn\\
&&\left. +4 k^2 r \omega ^2 (-72 r+17 i \omega )+128 r^2 \omega ^4\right)+12 \left(3 k^2-4 \omega ^2\right) \left(2 k^4+k^2 \left(-12 r^2-4 i r \omega \right.\right.\nn\\
&&\left.\left.\left.\left. +3 \omega ^2\right)+2 r \omega ^2 (8 r+3 i \omega )\right)\right)+2 \left(3 k^2-2 \omega ^2\right) \left(3 k^2-4 \omega ^2\right)^2 \left(k^2+2 i r \omega \right)\right)-3 \left(3 k^2 \right.\nn\\
&&\left.\left. -4 \omega ^2\right)^3 \left(k^2+2 i r \omega \right)\right\}+4 \zeta '(r) \left\{f(r) \left(k^2 f(r) \left(r^2 f(r) \left(-k^2 f(r) \left(5 k^4+3 k^2 r (-12 r \right.\right.\right.\right.\right.\nn\\
&&\left. +7 i \omega )+48 r^2 \omega ^2\right)+9 k^6+k^4 \left(-180 r^2+33 i r \omega -26 \omega ^2\right)+4 k^2 r \omega ^2 (96 r+5 i \omega ) \nn\\
&&\left. -192 r^2 \omega ^4\right)+3 k^8+4 k^6 \left(36 r^2+6 i r \omega -\omega ^2\right)+k^4 r \left(324 r^3+243 i r^2 \omega -324 r \omega ^2 \right.\nn\\
&&\left.\left. -32 i \omega ^3\right)-8 k^2 r^2 \omega ^2 \left(90 r^2+81 i r \omega -34 \omega ^2\right)+48 r^3 \omega ^4 (8 r+9 i \omega )\right)+\left(3 k^2-4 \omega ^2\right) \left(3 k^8 \right.\nn\\
&& -k^6 \left(63 r^2+4 \omega ^2\right)-k^4 r \left(108 r^3+171 i r^2 \omega -126 r \omega ^2+8 i \omega ^3\right)+8 k^2 r^2 \omega ^2 \left(18 r^2  \right.\nn \\
&&\left.\left.\left.\left.\left. +27 i r \omega-\omega ^2\right)-16 i r^3 \omega ^5\right)\right)+9 k^2 r^2 \left(3 k^2-4 \omega ^2\right)^2 \left(k^2+2 i r \omega \right)\right\}\right]\\
\mathcal{Q}_{so}^{(1)} & = & -\frac{1}{r \left(2 i r \omega + k^2\right) \left(f(r) k^2+3 k^2-4 \omega ^2\right)^3} \left[\left\{\omega  (3 r+2 i \omega ) \left(2 r \omega -i k^2\right) \left(3 k^2-4 \omega ^2\right)^3\right.\right.\nn \\
&& +f(r) \left(f(r) \left(f(r) \left(-3 k^8  +18 \left(-19 r^2-2 i \omega  r+\omega ^2\right) k^6+4 r (3 r-i \omega ) \left(108 r^2+99 i \omega  r \right.\right.\right.\right.\nn\\
&&\left. -26 \omega ^2\right) k^4-8 r^2 \omega ^2 \left(360 r^2+234 i \omega  r-85 \omega ^2\right) k^2 +32 r^3 \omega ^4 (48 r+35 i \omega )+f(r) \left(3 k^8 \right.\nn\\
&& +r (180 r+23 i \omega ) k^6-2 r^2 \left(576 r^2-108 i \omega  r+257 \omega ^2\right) k^4+64 r^3 \omega ^2 (30 r-i \omega ) k^2 \nn\\
&&\left.\left. -r^2 \left(43 k^4+6 r (41 i \omega -40 r) k^2+320 r^2 \omega ^2\right) f(r) k^2-512 r^4 \omega ^4\right)\right)-\left(3 k^2-4 \omega ^2\right) \left(9 k^6  \right.\nn \\
&&\left.\left.\left. +6 \left(18 r^2+3 i \omega  r-7 \omega ^2\right) k^4+8 i r \omega  \left(45 r^2+42 i \omega  r-11 \omega ^2\right) k^2+8 r^2 \omega ^3 (\omega -60 i r)\right)\right) k^2  \right.\nn \\
&&\left.\left. +\left(k^2+2 i r \omega \right) \left(3 k^2-4 \omega ^2\right)^2 \left(3 k^4+\left(9 r^2-18 i \omega  r+14 \omega ^2\right) k^2-4 i r \omega ^3\right)\right)\right\} \zeta ''(r) r^3 \nn\\
&& +\left\{\left(k^2+2 i r \omega \right) \left(3 k^6+2 \omega  (-3 i r-2 \omega ) k^4-6 r^2 \left(9 r^2+6 i \omega  r-2 \omega ^2\right) k^2  \right.\right.\nn \\
&&\left. +72 r^4 \omega ^2\right) \left(3 k^2-4 \omega ^2\right)^2+f(r) \left(2 \left(3 k^2-4 \omega ^2\right) \left(3 k^{10}+\left(63 r^2+18 i \omega  r-4 \omega ^2\right) k^8  \right.\right.\nn \\
&&\left. -r \left(189 r^3+18 i \omega  r^2+66 \omega ^2 r+28 i \omega ^3\right) k^6+2 r^2 \omega  \left(-297 i r^3+315 \omega  r^2+6 i \omega ^2 r+28 \omega ^3\right) k^4  \right.\nn \\
&&\left. +8 i r^3 \omega ^3 \left(63 r^2+18 i \omega  r+4 \omega ^2\right) k^2+32 r^4 \omega ^5 (6 i r+\omega )\right)+f(r) \left(3 k^{12}+\left(-90 r^2+36 i \omega  r  \right.\right.\nn \\
&&\left. -4 \omega ^2\right) k^{10}+12 r \left(171 r^3+63 i \omega  r^2+24 \omega ^2 r-4 i \omega ^3\right) k^8+4 r^2 \left(972 r^4+1971 i \omega  r^3   \right.\nn \\
&&\left. -1863 \omega ^2 r^2-186 i \omega ^3 r-32 \omega ^4\right) k^6-32 r^3 \omega ^2 \left(270 r^3+531 i \omega  r^2-243 \omega ^2 r-i \omega ^3\right) k^4   \nn \\
&& +64 r^4 \omega ^4 \left(72 r^2+132 i \omega  r-13 \omega ^2\right) k^2+2 r^2 f(r) \left(3 \left(-15 k^8-2 \left(186 r^2+37 i \omega  r-9 \omega ^2\right) k^6   \right.\right.\nn \\
&&\left. +2 r \left(-396 r^3-417 i \omega  r^2+339 \omega ^2 r+50 i \omega ^3\right) k^4+8 r^2 \omega ^2 \left(180 r^2+232 i \omega  r-73 \omega ^2\right) k^2   \right.\nn \\
&&\left. -64 r^3 (8 r+13 i \omega ) \omega ^4\right)+f(r) \left(-3 k^8+r (21 r-20 i \omega ) k^6+2 r^2 \left(684 r^2-24 i \omega  r+43 \omega ^2\right) k^4   \right.\nn \\
&&\left. -8 r^3 (300 r+91 i \omega ) \omega ^2 k^2+r^2 \left(47 k^4+12 r (17 i \omega -30 r) k^2+480 r^2 \omega ^2\right) f(r) k^2   \right.\nn \\
&&\hspace{1cm}\left.\left.\left.\left.\left.\left. +768 r^4 \omega ^4\right)\right) k^2+256 i r^5 \omega ^7\right)\right)\right\} \zeta '(r)\right]
\end{eqnarray}

After near horizon analysis of the master equations we find the first pole skipping point at $\omega_1=-\frac{3}{2}ir_0$ and
\begin{align}
& k_1^4+9 r_0^4-\lambda\frac{ \xi  \left(-k_1^6+9 (\xi +3) k_1^2 r_0^4+27 \xi r_0^6\right) \phi (r_0)^{p }}{r_0^2}\nn\\ 
& -\frac{\beta^2  \xi  \phi (r_0)^2 \left(k_1^6 (\xi +2 p )+3 k_1^4 r_0^2 (\xi -3p )-36 p  k_1^2 r_0^4-27 p r_0^6\right)}{p ^2 \left(k_1^2+3 r_0^2\right)}=0,    
\end{align}
Similarly one can find higher-order pole-skipping points. Since $\beta$ and $\lambda$ are two independent perturbative parameters, we can choose to observe the matter field's effect or the scalar-Gauss-Bonnet interaction's effect by switching the particular ranges of these parameters. For example, $\lambda\ll\beta^2\ll 1$ helps us to neglect the interaction effect and to consider the effect of matter stress-tensor only. However here we have considered the opposite one.

\bibliographystyle{jhep}
\bibliography{NPB}

  \end{document}